\documentclass[acmsmall,screen,nonacm]{acmart}
\usepackage{hyperref}
\usepackage{url}
\usepackage{graphicx} % Required for inserting images
\usepackage[utf8]{inputenc}
\usepackage[english]{babel}
\usepackage{amsmath}
\usepackage{amsfonts}
\usepackage{aliascnt}
\usepackage{listings}
\usepackage{amsthm}
\usepackage{bussproofs}
\usepackage{dirtytalk}
\usepackage{float}
\usepackage{xspace}
\usepackage{xifthen}
\usepackage{tikz}
\usepackage{stmaryrd}
\usepackage{framed}
\usetikzlibrary{positioning}

\usepackage[ruled,vlined,linesnumbered]{algorithm2e}

\SetAlFnt{\fontsize{8.5}{10}\selectfont}
\SetArgSty{textup}
\SetProgSty{textup}
\SetFuncArgSty{textup}
\SetFuncSty{textit}
\SetKwProg{procedure}{def}{}{}
\SetKwProg{datastructure}{Datastructure}{}{}
\SetKwSwitch{Switch}{Case}{Other}{match}{:}{case}{otherwise}{endcase}{endsw}

\SetKw{KwCont}{continue}
\SetKw{KwBreak}{break}

\newcommand{\cmnt}[1]{\tcp*[r]{\footnotesize #1}}

\newcommand{\lineElseIf}[2]{\textbf{else if}~{#1}~\textbf{then}~{#2}\;}

\SetKwComment{Comment}{/* }{ */}

\newcommand{\coqrefstyle}[1]{\textit{\texttt{\small\color{dkdkgreen} Rocq: \texttt{#1}}}}
\newcommand{\coqref}[1]{, \coqrefstyle{#1}}
\newcommand{\coqrefline}[1]{, \\\hspace*{2em}\coqrefstyle{#1}}

\newtheorem{theorem}{Theorem}[section]

\newaliascnt{lemma}{theorem}
\newtheorem{lemma}[lemma]{Lemma}
\aliascntresetthe{lemma}

\theoremstyle{definition}
\newaliascnt{definition}{theorem}
\newtheorem{definition}[definition]{Definition}
\aliascntresetthe{definition}
\newcommand{\simon}[1]{}
\newcommand{\viktor}[1]{}
\newcommand{\sankalp}[1]{}

\newcommand{\lOLP}{\ensuremath{\overline{\OLP}}\xspace}
\newcommand{\lBLP}{\ensuremath{\overline{\BLP}}\xspace}

\newcommand{\OLP}{\ensuremath{{\mathit{OL}^+}}\xspace}
\newcommand{\BLP}{\ensuremath{{\mathit{BL}^+}}\xspace}
\newcommand{\LP}{\ensuremath{{\mathit{L}^+}}\xspace}

\renewcommand{\L}{\mathit{L}\xspace}
\newcommand{\OL}{\mathit{OL}\xspace}

\newcommand{\SCP}{\ensuremath{{\mathit{SC}^+}}\xspace}

\newcommand{\CFP}{\ensuremath{{\mathit{CF}^+}}\xspace}
\newcommand{\CFPD}{\ensuremath{{\mathit{CF}^+_\delta}}\xspace}

\newcommand{\CFPBL}{\ensuremath{{\mathit{CF}^+_{BL}}}\xspace}

\newcommand{\A}{\mathcal A}
\newcommand{\B}{\mathcal B}

\newcommand{\F}{\mathcal F}

\renewcommand{\O}{\mathcal O}
\newcommand{\T}{\mathcal T}

\newcommand{\nle}{\not\leq}
\newcommand{\nge}{\not\geq}

\renewcommand{\slash}[1]{{{\kern-0.05em}/{\kern-0.08em}{#1}}}

\newcommand{\simw}[1]{\sim_{{\kern-0.1em}#1}}
\newcommand{\slashsimw}[1]{\slash{\simw{#1}}}
\newcommand{\term}[2]{\mathcal{T}_{#2}%
    \ifthenelse{\isempty{#1}}%
    {}%
    {(#1)}%
}

\newcommand{\eqclass}[2]{{[#1]}_{#2}}

\newcommand{\kappaol}{\kappa_{F_{\OL}}{\kern-0.2em}}
\newcommand{\kappal}{\kappa_{F_{L}}{\kern-0.2em}}
\newcommand{\kappax}[1]{\kappa_{F_{#1}}{\kern-0.2em}}

\newcommand{\xyz}{x_1, ..., y_1, ..., z_1, ...}
\newcommand{\Sxyz}{S_{x_1}, ..., S_{y_1}, ..., S_{z_1}, ...}
\newcommand{\Txyz}{T_{x_1}, ..., T_{y_1}, ..., T_{z_1}, ...}

\newcommand{\scrule}[1]{\textsc{#1}}

\newcommand{\clausesS}[1]{\textsf{clauses}({#1})}
\newcommand{\sizeOf}[1]{\|{#1}\|}

\newcommand{\FOLm}{\mathit{F{\kern-0.1em}({\kern-0.1em}O{\kern-0.1em}L{\kern-0.1em})\textsuperscript{2}}}

\newcommand{\dashvdash}{\dashv\vdash}

\newcommand{\rocq}{Rocq}

\usepackage{listings}

\definecolor{keywordcolor}{rgb}{0.7, 0.1, 0.1}   % red
\definecolor{tacticcolor}{rgb}{0.0, 0.1, 0.6}    % blue
\definecolor{commentcolor}{rgb}{0.4, 0.4, 0.4}   % grey
\definecolor{symbolcolor}{rgb}{0.0, 0.1, 0.6}    % blue
\definecolor{sortcolor}{rgb}{0.1, 0.5, 0.1}      % green
\definecolor{attributecolor}{rgb}{0.7, 0.1, 0.1} % red

\definecolor{rosishlightgray}{rgb}{0.96, 0.94, 0.92}
\definecolor{bluishlightgray}{rgb}{0.94, 0.96, 0.98}
\definecolor{greenishlightgray}{rgb}{0.94, 0.98, 0.96}
\definecolor{lightgreen}{rgb}{0, 0.65, 0}

\definecolor{dkred}{rgb}{0.4,0,0}
\definecolor{dkgreen}{rgb}{0,0.4,0}
\definecolor{dkdkgreen}{rgb}{0,0.3,0}
\definecolor{dkblue}{rgb}{0,0,0.4}
\definecolor{dkviolet}{rgb}{0.3,0,0.5}
\definecolor{green}{rgb}{0, 0.6, 0}

\lstdefinelanguage{pseudo}{
    alsoletter={@,=,>},
    keywordstyle = {\color{blue}},
    keywordstyle = [2]{\color{blue}},
    commentstyle = \color{dkgreen},
    backgroundcolor = \color{rosishlightgray},
    morekeywords = [2]{abstract, case, class, def, do, Input, Output, then,
        else, extends, false, free, if, implicit, match,
        object, true, val, var, while, sealed, or,
        for, dependent, null, type, with, try, catch, finally,
        import, final, return, new, override, this, trait,
        private, public, protected, package, throw, let, rec, undefined, NaN},
    sensitive = true, 
    numbers=left,
    stepnumber=1,
    sensitive = true, 
    numbers=none, % now none is default, use [numbers=left] if needed with in lstlisting
    morecomment = [l]{//},
    morecomment = [s]{/*}{*/},
    morestring = [b]",  
    otherkeywords = {},
    mathescape = true,
    escapeinside = {{*@}{@*}},
    literate = {
        {λ}{\(\lambda\)}{1}
    },
}

\lstdefinelanguage{scala}{
    alsoletter={@,=,>},
    keywordstyle = {\color{blue}},
    keywordstyle = [2]{\color{blue}},
    commentstyle = \color{commentcolor},
    backgroundcolor = \color{rosishlightgray},
    morekeywords = [2]{abstract, case, class, def, do, Input, Output, then,
        else, extends, false, free, if, implicit, match,
        object, true, val, var, while, sealed, or,
        for, dependent, null, type, with, try, catch, finally,
        import, final, return, new, override, this, trait,
        private, public, protected, package, throw},
    sensitive = true, 
    numbers=left,
    stepnumber=1,
    sensitive = true, 
    numbers=none, % now none is default, use [numbers=left] if needed with in lstlisting
    morecomment = [l]{//},
    morecomment = [s]{/*}{*/},
    morestring = [b]",  
    otherkeywords = {},
    mathescape = true,
    escapeinside = {{*@}{@*}},
    literate = {
        {λ}{\(\lambda\)}{1}
    },
}

\lstdefinelanguage{java}{
    alsoletter={@,=,>},
    keywordstyle = {\color{blue}},
    keywordstyle = [2]{\color{blue}},
    commentstyle = \color{commentcolor},
    backgroundcolor = \color{greenishlightgray},
    morekeywords = [2]{abstract, class, implements, interface, def, do, Input, Output, then,
        else, extends, false, free, if, implicit, match,
        object, true, val, var, while, sealed, or,
        for, dependent, null, type, with, try, catch, finally,
        import, final, return, new, override, this, trait,
        private, public, protected, package, throw},
    sensitive = true, 
    numbers=left,
    stepnumber=1,
    sensitive = true, 
    numbers=none, % now none is default, use [numbers=left] if needed with in lstlisting
    morecomment = [l]{//},
    morecomment = [s]{/*}{*/},
    morestring = [b]",  
    otherkeywords = {;+},
    mathescape = true,
    escapeinside = {{*@}{@*}},
}

\lstdefinelanguage{lisa}{
    keepspaces=true,
    extendedchars = true,
    inputencoding = utf8,
    alsoletter={@,=,>},
    backgroundcolor = \color{rosishlightgray},
    keywordstyle = {\color{blue}},
    keywordstyle = [2]{\color{blue}},
    keywordstyle = [3]{\color{green}},
    keywordstyle = [4]{\color{dkviolet}},
    commentstyle = \color{comments},
    basicstyle=\footnotesize\ttfamily,
    morekeywords = [2]{abstract, case, class, def, do, Input, Output, then,
        else, extends, false, free, if, implicit, match,
        object, true, val, var, while, sealed, or,
        for, dependent, null, type, with, try, catch, finally,
        import, final, return, new, override, this, trait,
        private, public, protected, package, throw, given},
    morekeywords = [3]{have, andThen, thenHave, Theorem, by, DEF, The, Lemma, subproof, assume, Case},
    morekeywords = [4]{BETA, INST, TRANS},
    sensitive = true, 
    numbers=none, % now none is default, use [numbers=left] if needed with in lstlisting
    stepnumber=1,
    morecomment = [l]{//},
    morecomment = [s]{/*}{*/},
    morestring = [b]",  
    otherkeywords = {;,<<,>>,++},
}

\lstdefinelanguage{tptp}{
    keepspaces=true,
    extendedchars = true,
    inputencoding = utf8,
    backgroundcolor = \color{bluishlightgray},
    keywordstyle = {\color{blue}},
    keywordstyle = [2]{\color{blue}},
    keywordstyle = [3]{\color{dkgreen}},
    keywordstyle = [4]{\color{dkviolet}},
    commentstyle = \color{comments},
    basicstyle={\footnotesize\ttfamily},
    morekeywords = [2]{hyp, rightImplies, rightForall, rightExists},
    morekeywords = [3]{inference, status},
    morekeywords = [4]{fof, fot, let, plain, assumption, simplify, axiom, conjecture},
    sensitive = true, 
    numbers=none,
    stepnumber=1,
    morecomment = [l]{//},
    morecomment = [s]{/*}{*/},
    morestring = [b]",
    literate=
    {==>}{{$\Longrightarrow\;$}}1
    {=>}{{$\Rightarrow\;$}}1
    {-->}{{$\longrightarrow\;$}}1,
}

\lstdefinelanguage{Coq}{ 
    mathescape=true,
    texcl=false, 
    morekeywords=[1]{Section, Module, End, Require, Import, Export,
        Variable, Variables, Parameter, Parameters, Axiom, Hypothesis,
        Hypotheses, Notation, Local, Tactic, Reserved, Scope, Open, Close,
        Bind, Delimit, Definition, Let, Ltac, Fixpoint, CoFixpoint, Add,
        Morphism, Relation, Implicit, Arguments, Unset, Contextual,
        Strict, Prenex, Implicits, Inductive, CoInductive, Record,
        Structure, Canonical, Coercion, Context, Class, Global, Instance,
        Program, Infix, Theorem, Lemma, Corollary, Proposition, Fact,
        Remark, Example, Proof, Goal, Save, Qed, Defined, Hint, Resolve,
        Rewrite, View, Search, Show, Print, Printing, All, Eval, Check,
        Projections, inside, outside, Def},
    morekeywords=[2]{forall, exists, exists2, fun, fix, cofix, struct,
        match, with, end, as, in, return, let, if, is, then, else, for, of,
        nosimpl, when},
    morekeywords=[3]{Type, Prop, Set, true, false, option},
    morekeywords=[4]{pose, set, move, case, elim, apply, clear, hnf,
        intro, intros, generalize, rename, pattern, after, destruct,
        induction, using, refine, inversion, injection, rewrite, congr,
        unlock, compute, ring, field, fourier, replace, fold, unfold,
        change, cutrewrite, simpl, have, suff, wlog, suffices, without,
        loss, nat_norm, assert, cut, trivial, revert, bool_congr, nat_congr,
        symmetry, transitivity, auto, split, left, right, autorewrite},
    morekeywords=[5]{by, done, exact, reflexivity, tauto, romega, omega,
        assumption, solve, contradiction, discriminate},
    morekeywords=[6]{do, last, first, try, idtac, repeat},
    morecomment=[s]{(*}{*)},
    showstringspaces=false,
    morestring=[b]",
    morestring=[d]’,
    tabsize=3,
    extendedchars=false,
    sensitive=true,
    breaklines=false,
    basicstyle=\small,
    captionpos=b,
    columns=[l]flexible,
    identifierstyle={\ttfamily\color{black}},
    keywordstyle=[1]{\ttfamily\color{violet}},
    keywordstyle=[2]{\ttfamily\color{green}},
    keywordstyle=[3]{\ttfamily\color{blue}},
    keywordstyle=[4]{\ttfamily\color{blue}},
    keywordstyle=[5]{\ttfamily\color{red}},
    stringstyle=\ttfamily,
    commentstyle={\ttfamily\color{dkgreen}},
    literate=
    {\\forall}{{\color{dkgreen}{$\forall\;$}}}1
    {\\exists}{{$\exists\;$}}1
    {<-}{{$\leftarrow\;$}}1
    {=>}{{$\Rightarrow\;$}}1
    {==}{{\code{==}\;}}1
    {==>}{{\code{==>}\;}}1
    {->}{{$\rightarrow\;$}}1
    {<->}{{$\leftrightarrow\;$}}1
    {<==}{{$\leq\;$}}1
    {\#}{{$^\star$}}1 
    {\\o}{{$\circ\;$}}1 
    {\@}{{$\cdot$}}1 
    {\/\\}{{$\wedge\;$}}1
    {\\\/}{{$\vee\;$}}1
    {++}{{\code{++}}}1
    {~}{{$\sim$}}1
    {\@\@}{{$@$}}1
    {\\mapsto}{{$\mapsto\;$}}1
    {\\hline}{{\rule{\linewidth}{0.5pt}}}1
}[keywords,comments,strings]

\lstdefinelanguage{lean}{
mathescape=false,
texcl=false,
morekeywords=[1]{
import, prelude, protected, private, noncomputable, definition, meta, renaming,
hiding, parameter, parameters, begin, constant, constants,
lemma, variable, variables, theory,
print, theorem, example,
open, as, export, override, axiom, axioms, inductive, with,
structure, record, universe, universes,
alias, help, precedence, reserve, declare_trace, add_key_equivalence,
match, infix, infixl, infixr, notation, postfix, prefix, instance,
eval, reduce, check, end, this,
using, using_well_founded, namespace, section,
attribute, local, set_option, extends, include, omit, class,
raw, replacing,
calc, have, show, suffices, by, in, at, let, forall, Pi, fun,
exists, if, dif, then, else, assume, obtain, from, register_simp_ext, unless, break, continue,
mutual, do, def, run_cmd, const,
partial, mut, where, macro, syntax, deriving,
return, try, catch, for, macro_rules, declare_syntax_cat, abbrev},
morekeywords=[2]{Sort, Type, Prop},
morekeywords=[3]{
assumption,
apply, intro, intros, allGoals,
generalize, clear, revert, done, exact,
refine, repeat, cases, rewrite, rw,
simp, simp_all, contradiction,
constructor, injection,
induction,
},
literate=
{α}{{\ensuremath{\mathrm{\alpha}}}}1
{β}{{\ensuremath{\mathrm{\beta}}}}1
{γ}{{\ensuremath{\mathrm{\gamma}}}}1
{δ}{{\ensuremath{\mathrm{\delta}}}}1
{ε}{{\ensuremath{\mathrm{\varepsilon}}}}1
{ζ}{{\ensuremath{\mathrm{\zeta}}}}1
{η}{{\ensuremath{\mathrm{\eta}}}}1
{θ}{{\ensuremath{\mathrm{\theta}}}}1
{ι}{{\ensuremath{\mathrm{\iota}}}}1
{κ}{{\ensuremath{\mathrm{\kappa}}}}1
{μ}{{\ensuremath{\mathrm{\mu}}}}1
{ν}{{\ensuremath{\mathrm{\nu}}}}1
{ξ}{{\ensuremath{\mathrm{\xi}}}}1
{π}{{\ensuremath{\mathrm{\mathnormal{\pi}}}}}1
{ρ}{{\ensuremath{\mathrm{\rho}}}}1
{σ}{{\ensuremath{\mathrm{\sigma}}}}1
{τ}{{\ensuremath{\mathrm{\tau}}}}1
{φ}{{\ensuremath{\mathrm{\varphi}}}}1
{χ}{{\ensuremath{\mathrm{\chi}}}}1
{ψ}{{\ensuremath{\mathrm{\psi}}}}1
{ω}{{\ensuremath{\mathrm{\omega}}}}1
{Γ}{{\ensuremath{\mathrm{\Gamma}}}}1
{Δ}{{\ensuremath{\mathrm{\Delta}}}}1
{Θ}{{\ensuremath{\mathrm{\Theta}}}}1
{Λ}{{\ensuremath{\mathrm{\Lambda}}}}1
{Σ}{{\ensuremath{\mathrm{\Sigma}}}}1
{Φ}{{\ensuremath{\mathrm{\Phi}}}}1
{Ξ}{{\ensuremath{\mathrm{\Xi}}}}1
{Ψ}{{\ensuremath{\mathrm{\Psi}}}}1
{Ω}{{\ensuremath{\mathrm{\Omega}}}}1
{ℵ}{{\ensuremath{\aleph}}}1
{≤}{{\ensuremath{\leq}}}1
{≥}{{\ensuremath{\geq}}}1
{≠}{{\ensuremath{\neq}}}1
{≈}{{\ensuremath{\approx}}}1
{≡}{{\ensuremath{\equiv}}}1
{≃}{{\ensuremath{\simeq}}}1
{≤}{{\ensuremath{\leq}}}1
{≥}{{\ensuremath{\geq}}}1
{∂}{{\ensuremath{\partial}}}1
{∆}{{\ensuremath{\triangle}}}1 % or \laplace?
{∫}{{\ensuremath{\int}}}1
{∑}{{\ensuremath{\mathrm{\Sigma}}}}1
{Π}{{\ensuremath{\mathrm{\Pi}}}}1
{⊥}{{\ensuremath{\perp}}}1
{∞}{{\ensuremath{\infty}}}1
{∂}{{\ensuremath{\partial}}}1
{∓}{{\ensuremath{\mp}}}1
{±}{{\ensuremath{\pm}}}1
{×}{{\ensuremath{\times}}}1
{⊕}{{\ensuremath{\oplus}}}1
{⊗}{{\ensuremath{\otimes}}}1
{⊞}{{\ensuremath{\boxplus}}}1
{∇}{{\ensuremath{\nabla}}}1
{√}{{\ensuremath{\sqrt}}}1
{⬝}{{\ensuremath{\cdot}}}1
{•}{{\ensuremath{\cdot}}}1
{∘}{{\ensuremath{\circ}}}1
{⁻}{{\ensuremath{^{-}}}}1
{▸}{{\ensuremath{\blacktriangleright}}}1
{∧}{{\ensuremath{\wedge}}}1
{∨}{{\ensuremath{\vee}}}1
{¬}{{\ensuremath{\neg}}}1
{⊢}{{\ensuremath{\vdash}}}1
{⟨}{{\ensuremath{\langle}}}1
{⟩}{{\ensuremath{\rangle}}}1
{↦}{{\ensuremath{\mapsto}}}1
{←}{{\ensuremath{\leftarrow}}}1
{<-}{{\ensuremath{\leftarrow}}}1
{→}{{\ensuremath{\rightarrow}}}1
{↔}{{\ensuremath{\leftrightarrow}}}1
{⇒}{{\ensuremath{\Rightarrow}}}1
{⟹}{{\ensuremath{\Longrightarrow}}}1
{⇐}{{\ensuremath{\Leftarrow}}}1
{⟸}{{\ensuremath{\Longleftarrow}}}1
{∩}{{\ensuremath{\cap}}}1
{∪}{{\ensuremath{\cup}}}1
{⊂}{{\ensuremath{\subseteq}}}1
{⊆}{{\ensuremath{\subseteq}}}1
{⊄}{{\ensuremath{\nsubseteq}}}1
{⊈}{{\ensuremath{\nsubseteq}}}1
{⊃}{{\ensuremath{\supseteq}}}1
{⊇}{{\ensuremath{\supseteq}}}1
{⊅}{{\ensuremath{\nsupseteq}}}1
{⊉}{{\ensuremath{\nsupseteq}}}1
{∈}{{\ensuremath{\in}}}1
{∉}{{\ensuremath{\notin}}}1
{∋}{{\ensuremath{\ni}}}1
{∌}{{\ensuremath{\notni}}}1
{∅}{{\ensuremath{\emptyset}}}1
{∖}{{\ensuremath{\setminus}}}1
{†}{{\ensuremath{\dag}}}1
{ℕ}{{\ensuremath{\mathbb{N}}}}1
{ℤ}{{\ensuremath{\mathbb{Z}}}}1
{ℝ}{{\ensuremath{\mathbb{R}}}}1
{ℚ}{{\ensuremath{\mathbb{Q}}}}1
{ℂ}{{\ensuremath{\mathbb{C}}}}1
{⌞}{{\ensuremath{\llcorner}}}1
{⌟}{{\ensuremath{\lrcorner}}}1
{⦃}{{\ensuremath{\{\!|}}}1
{⦄}{{\ensuremath{|\!\}}}}1
{‖}{{\ensuremath{\|}}}1
{₁}{{\ensuremath{_1}}}1
{₂}{{\ensuremath{_2}}}1
{₃}{{\ensuremath{_3}}}1
{₄}{{\ensuremath{_4}}}1
{₅}{{\ensuremath{_5}}}1
{₆}{{\ensuremath{_6}}}1
{₇}{{\ensuremath{_7}}}1
{₈}{{\ensuremath{_8}}}1
{₉}{{\ensuremath{_9}}}1
{₀}{{\ensuremath{_0}}}1
{ᵢ}{{\ensuremath{_i}}}1
{ⱼ}{{\ensuremath{_j}}}1
{ₐ}{{\ensuremath{_a}}}1
{¹}{{\ensuremath{^1}}}1
{ₙ}{{\ensuremath{_n}}}1
{ₘ}{{\ensuremath{_m}}}1
{ₚ}{{\ensuremath{_p}}}1
{↑}{{\ensuremath{\uparrow}}}1
{↓}{{\ensuremath{\downarrow}}}1
{...}{{\ensuremath{\ldots}}}1
{·}{{\ensuremath{\cdot}}}1
{▸}{{\ensuremath{\triangleright}}}1
{Σ}{{\color{symbolcolor}\ensuremath{\Sigma}}}1
{Π}{{\color{symbolcolor}\ensuremath{\Pi}}}1
{∀}{{\color{symbolcolor}\ensuremath{\forall}}}1
{∃}{{\color{symbolcolor}\ensuremath{\exists}}}1
{λ}{{\color{symbolcolor}\ensuremath{\mathrm{\lambda}}}}1
{\$}{{\color{symbolcolor}\$}}1
{:=}{{\color{symbolcolor}:=}}1
{=}{{\color{symbolcolor}=}}1
{<|>}{{\color{symbolcolor}<|>}}1
{<\$>}{{\color{symbolcolor}<\$>}}1
{+}{{\color{symbolcolor}+}}1
{*}{{\color{symbolcolor}*}}1,
morecomment=[s][\color{commentcolor}]{/-}{-/},
morecomment=[l][\itshape \color{commentcolor}]{--},
showstringspaces=false,
keepspaces=true,
morestring=[b]",
morestring=[d],
tabsize=3,
extendedchars=false,
sensitive=true,
breaklines=true,
breakatwhitespace=true,
basicstyle=\ttfamily\small,
captionpos=b,
columns=[l]fullflexible,
identifierstyle={\ttfamily\color{black}},
keywordstyle=[1]{\ttfamily\color{keywordcolor}},
keywordstyle=[2]{\ttfamily\color{sortcolor}},
keywordstyle=[3]{\ttfamily\color{tacticcolor}},
keywordstyle=[4]{\ttfamily\color{attributecolor}},
stringstyle=\ttfamily,
commentstyle={\ttfamily\footnotesize },
}

\title{Orthologic Type Systems}
\author{Simon Guilloud and Viktor Kun\v{c}ak}
\date{}



\begin{document}

\begin{abstract}
We propose to use orthologic as the basis for designing type systems supporting intersection, union, and negation types in the presence of subtyping assumptions. We show how to extend orthologic to support monotonic and antimonotonic functions, supporting the use of type constructors in such type systems. We present a proof system for orthologic with function symbols, showing that it admits partial cut elimination. Using these insights, we present an $\mathcal O(n^2(1+m))$ algorithm for deciding the subtyping relation under $m$ assumptions. We also show $O(n^2)$ polynomial-time normalization algorithm, allowing simplification of types to their minimal canonical form.
\end{abstract}

\maketitle

\section{Introduction}

Partial orders are a basis for designing many type systems, program analysis and verification algorithms. Among partial orders, lattices stand out thanks to their rich set of operations and polynomial-time algorithms, both in the free lattice case \cite{whitmanFreeLattices1941} and for lattices modeling purely structural subtyping \cite[Section 4]{DBLP:conf/lics/Tiuryn92}. 
Models of types based on ideals define intersections and union types using set theoretic intersections and unions \cite[Page 117]{DBLP:journals/iandc/MacQueenPS86}.
In the last decade, type systems with intersections and unions have become more widely used.
Scala 3 introduced explicit unions and intersections, improving type inference and expressiveness compared to Scala 2. Intersections and union types are included in the formalization of Scala 3 type system aspects \cite{DBLP:journals/pacmpl/GiarrussoSTBK20} based on logical relations in the Iris framework \cite{DBLP:journals/jfp/JungKJBBD18}.  TypeScript \cite{DBLP:conf/popl/RastogiSFBV15} and Flow \cite{flow} also support intersection types. PHP from 8.1 \cite{php81_intersection_types},
also has limited support for unions and intersections, though not their general combination. 

These uses suggest that such systems are a natural fit for languages that were designed as dynamically typed.
Indeed, intersections naturally model combinations of records with different fields, whereas unions model a rich set of possible values that a function in a dynamically typed language can return. In the past, such observations have led to the line of soft typing for Scheme-like languages, leading to the analyses based on expressive type constraints \cite{DBLP:conf/popl/AikenWL94}.

Today, widely used implementations of these features are often incomplete, inefficient, or rely on unintuitive assumptions compared to expected lattice behaviours.
This is unfortunate since good algorithms for problems on lattices do exist \cite{whitmanFreeLattices1941}. 
Lattices are also the basis of abstract interpretation \cite{DBLP:conf/popl/CousotC79} and program analysis in general \cite{DBLP:books/daglib/0098888}, owing to their good theoretical properties. As an example, a classic abstract domain of integer intervals \cite[Section 7]{DBLP:conf/popl/CousotC77} with $\land$ defined as intersection and $[a_1,b_1] \lor [a_2,b_2] = [\min(a_1,a_2), \max(b_1,b_2)]$ is a useful lattice that is non-distributive, as we can see by considering, e.g. $[1,3] \land ([0,2] \lor [4,5]) = [1,3]$, whereas 
$
([1,3] \land [0,2]) \lor ([1,3] \land [4,5])\   = \ [1,2]
$.

In addition to unions and intersections, negations of types are also a natural notion. Their systematic implementation was pioneered in languages such as CDuce \cite{castagnaProgrammingUnionIntersection2024} and MLscript \cite{parreauxMLstructPrincipalType2022}. They are also a popular request for TypeScript \cite{negatedTypesTS}. Another natural fit for 
intersection, union, and negation of types are expressive predicate refinement type systems, where propositional logic operations naturally lead to the corresponding operations on subtypes. Such systems have been formalized by interpreting types as sets of terms \cite{DBLP:journals/pacmpl/HamzaVK19}.

\begin{table*}[bth]    
    \centering
    \begin{tabular}{r c @{\hskip 2em} | @{\hskip 2em} r c}
         V1: & $x \lor y = y \lor x$  & V1': & $x \land y = y \land x$ \\
         V2: & $x \lor ( y \lor z) = (x \lor y) \lor z$  & V2': & $x \land ( y \land z) = (x \land y) \land z$ \\
         V3: & $x \lor x = x$  & V3': & $x \land x = x$ \\
         V4: & $x \lor (x \land y) = x$ & V4': & $x \land (x \lor y) = x$   \\
         V5: & $x \lor \top = \top$  & V5': & $x \land \bot = \bot$ \\
         V6: & $x \lor \bot = x$  & V6': & $x \land \top = x$ \\
         V7: & $\neg \neg x = x$\\
         V8: & $x \lor \neg x = \top$  & V8': & $x \land \neg x = \bot$ \\
         V9: & $\neg (x \lor y) = \neg x \land \neg y$  & V9': &  $\neg (x \land y) = \neg x \lor \neg y$ \\
    \end{tabular}
    \vspace{2ex}
    \
    \caption{Laws of ortholattices, algebraic varieties with signature $(S, \land, \lor, \bot, \top, \neg)$.}
    \label{tab:algebraiclaws}
\end{table*}

In summary, a structure with unions, intersections, and complements is of great potential value. These operations form the signature of a Boolean algebra. Unfortunately, entailment in Boolean algebras is coNP-complete, with normal forms exponentially larger than the starting term in the general case.
In languages that support distributivity of unions and intersections, this leads to exponential time type checking (see \autoref{sec:typelattices}). 

As an alternative to Boolean algebra, our paper follows an approach based on ortholattices, whose good algorithmic properties were recently established and formalized \cite{guilloudFormulaNormalizationsVerification2023, guilloudOrthologicAxioms2024, guilloudVerifiedOptimizedImplementation2025}. 
\autoref{tab:algebraiclaws} shows the laws of ortholattices.
Ortholattices were shown to have quadratic time normalization and entailment procedures, or cubic time in the presence of background constraints (axioms) \cite{guilloudOrthologicAxioms2024}. These structures do not satisfy, in general, the distributivity rule, but neither do all type lattices in programming languages deployed in practice, such as Flow \cite{flow}, or studied in the literature \cite{jiangBidirectionalHigherRankPolymorphism2025}. 

Despite good algorithmic properties of ortholattices, a type system based on these structures alone can only accounts for behavior of base types, which makes their use limited, especially for programs using collections and first-class functions. A key idea of this paper is hence the following: model type constructors as (possibly monotonic or antimonotonic) functions on ortholattices.
We therefore generalize the theory ortholattices and their efficient algorithms to support such functions, enabling the modelling of function types, pairs, records, algebraic datatypes, and classes. We name the resulting algebraic class  of ortholattices (resp.~bounded lattices) with monotonic and antimonotonic function symbols {\OLP} (resp.~\BLP). 
\begin{figure*}[bth]    
\[
\frac{x \le x', \ \ y \le y'}{p(x,y) \le p(x', y')}
\qquad\qquad
\frac{x' \le x, \ \ y \le y'}{f(x,y) \le f(x', y')}
\]
    \caption{Examples of Functions on Lattices with Monotonicity Properties}
    \label{tab:exampleMonotonicity}
\end{figure*}

\autoref{tab:exampleMonotonicity} illustrates the monotonicity laws, where $p$ might represent a covariant pair constructor, whereas $f$ represents a function type constructor contravariant in the first and covariant in the second argument. In general, a function may be monotonic in some, antimonotonic in others, and non-monotonic in yet other arguments. By supporting lattices on types that include constructors, our framework bears similarity to those studied for structural and non-structural subtyping \cite{suFirstorderTheorySubtyping2002}. This area includes some long-standing open problems, including a subtyping Problem number 16 in the TLCA list of open problems \cite{TLCAListOpen}, shown to be PSPACE hard \cite{DBLP:conf/icalp/HengleinR98}. However, our axiomatization of types differs from such models because we only assume monotonicity, without restricting the meaning of types to trees with a fixed subtyping relation. In the tree model, despite non-structural subtyping, charactizations such as \cite[Theorem 2.1]{DBLP:journals/iandc/Pottier01} imply, e.g.,
\begin{equation} \label{eq:syntacticFun}
   (x_1 \to x_2) \land (y_1 \to y_2) \ \ = \ \ (x_1 \lor y_1) \to (x_2 \land y_2)
\end{equation}
Many intersection type systems with greater expressive power do not satisfy \eqref{eq:syntacticFun} and contain counterexample to it \cite{DBLP:journals/ndjfl/CoppoD80, DBLP:conf/popl/DunfieldP04, jiangBidirectionalHigherRankPolymorphism2025, castagnaProgrammingUnionIntersection2024}. 
In contrast to \eqref{eq:syntacticFun}, monotonicity properties that we support continue to hold in the expressive intersection type systems. With our model we therefore obtain wider applicability of the algorithms.

Crucially, we obtain polynomial-time algorithms for subtyping and normalization.  In fact, we show that the complexity of normalization, as well as entailment in the presence of axioms remains the same as in the case of pure ortholattices. This yields new algorithms that can be used in type systems, compositional program analyses, as well as verification. 

Our paper focuses on the algorithmic problem of deciding subtyping relations in the presence of union, intersection, negation types, (anti)monotonic type constructors and subtyping constraints. We leave the broader aspects of designing type systems and corresponding term language based on these principles to subsequent work, contributing instead to the toolbox of algorithms for type system design.
%
% This is too vague and comes across as non-factual self-praising
%Hence, our key contribution is a new approach to type systems with subtyping that is sound, complete, and efficient.
%, simple, reliable, generally applicable and easy to reason with}.
%\footnote{In particular, our algorithm may seem less complex, and hence possibly less impressive than other approaches in the literature; this is a feature}.

In addition to subtyping, we consider the problem of simplification of types, which has significant practical interest \cite{pottierSimplifyingSubtypingConstraints1996}. Many existing compilers, such as for Scala, Flow and Typescript, perform 
simplifications on type expressions. Among the motivations are better error messages and improved efficiency, in particular in the presence of union and intersection types. We present a quadratic-time algorithm that computes a normal form (in the absence of axioms), with a remarkable property that the computed the normal form is always the smallest type equivalent to the input.

\paragraph{Contributions}
This paper makes the following contributions:

\begin{enumerate}
%\item In \autoref{sec:preliminaries}, we introduce \OLP, the variety of ortholattices with monotonic function and \BLP, the variety of lattices with monotonic functions. We also present relevant notation and definitions from universal algebra. We then discuss in detail union (\lstinline{|}), intersection (\lstinline{|}) and negation (\lstinline{|}) types, and their interest and usage when types are embedded in an (ortho)lattice.
    \item We propose the use of free ortholattices as a theoretically well founded and algorithmically efficient approach to subtyping. Such an approach offer union, intersection and negation types, as well as type constructors with covariant and contravariant parameters and subtyping constraints. Thanks to the results below, this system admits polynomial-time subtyping and normalization algorithms. 
    Considering types as a \textit{free} (ortho)lattice can be thought of as a \textit{maximally} non-structural type system, offering safety even under an open-world assumption.
    
    \item  We present a sound and complete proof system for \OLP: orthologic extended with arbitrary monotonic and antimonotonic function symbols (which can model type constructors). As a key result, we show (in \autoref{sec:generalizedwordproblem}) partial cut (i.e. transitivity) elimination for this system. We formalize this result in \rocq.
    
    \item We use the above cut elimination theorem to develop an algorithm (\autoref{sec:subalgorithm}) for checking inequalities in $\OLP$ in the presence of axioms, with time complexity $\O(n^2(1+|A|))$, where $m$ is the number of axioms. We introduce a new approach for formulating orthologic algorithms, via reduction to a polynomial number of propositional Horn clauses, which stands in contrast to previous error-prone formulation based on memoization.
    
% make this more precise
    \item We present (\autoref{sec:normalization}) a characterization of minimal form for bounded lattices with monotonic functions, \BLP. We use this characterization to show an $\O(n^2)$-time normalization algorithm. We then solve the problem of normalization for \OLP \textit{up to equivalence in \BLP}. We combine this procedure with normalization in lattices to obtain a complete normalization algorithm, which can be used to efficiently compute a canonical and minimal form for types. By working with a concrete representation of the free lattices and ortholattices with function symbols, we avoid the need for isomorphism theorems or category theory present in some of the past work \cite{guilloudEquivalenceCheckingOrthocomplemented2022, dolanAlgebraicSubtypingDistinguished2017}, making our proofs self-contained.
% \item In \autoref{sec:records} and \autoref{sec:classes}, we illustrate possible uses of our systems with examples of modeling  inheritance, record types, monomorphic isorecursive types, F-bounded polymorphism, and pattern matching with exhaustivity checks.
%\item Finally, we discuss additional related research in \autoref{sec:relatedwork}
\end{enumerate}

\section{Union, Intersection, and Negation Types in Type Checkers}
\label{sec:typelattices}

This section illustrates informally how reasoning about intersections, unions, and negation arises in realistic programming language type systems. We expect that our algorithm can contribute to making these type systems more predictable and efficient. We discuss the behavior and limitations of unions and intersections as implemented in some popular programming languages. We then discuss the potential for negation types, and outline how to model additional typing constructs in a system based on free ortholattices with function symbols. Our main technical results are in subsequent sections.

\subsection{Unions, Intersections, and the Price of Distributivity}

We highlight three programming languages that were deployed on large code bases and that support subtyping with union and intersection types: Scala, TypeScript and Flow. Following common ASCII notation, we use {\tt \&} to denote intersection ($\land$) and {\tt |} to denote union $(\lor)$.

\paragraph{Scala}
Scala 3 supports union and intersection types with the laws of a lattice. Laws V1-V4 can be proven using the type checker, as can their duals, V1'-V4': given unconstrained types {\tt A}, {\tt B}, {\tt C}, Scala version 3.6.3 type checker accepts all of the following:
\begin{lstlisting}
summon[(A|B) =:= (B|A)]               summon[(A & B) =:= (B & A)]           
summon[(A|A) =:= A]                   summon[(A & A) =:= A]  
summon[(A|(B|C)) =:= ((A|B)|C)]       summon[(A & (B & C)) =:= ((A & B) & C)]
summon[(A|(A&B)) =:= A]               summon[(A & (A|B)) =:= A]
\end{lstlisting}
Here {\tt LHS =:= RHS} is a Scala expression that invokes two subtyping checks between LHS and RHS, succeeding (with an evidence object) if they both succeed. Furthermore, using Scala's {\tt Nothing} for the least element $\bot$ and {\tt Any} for $\top$, the equations V5-V6 and V5'-V6' also hold:
\begin{lstlisting}
summon[(A | Any) =:= Any]             summon[(A & Nothing) =:= Nothing]
summon[(A | Nothing) =:= A]           summon[(A & Any) =:= A]
\end{lstlisting}

Note that, in some cases, Scala type system does not reason completely with assumptions, missing an opportunity to apply the transitivity of subtyping as well:
\begin{lstlisting}
type A
type C
def foo[B >: A <: C](x: A) : C = x     // does not compile
def bar[B >: A <: C](x: A) : C = (x:B) // compiles, thanks to the x:B hint
\end{lstlisting}
Our algorithm supports reasoning with assumptions (axioms). Maintaining a global store of such assumptions and using our algorithm would eliminate such source of incompleteness.

Going beyond the axioms of lattices, Scala also satisfies distributivity laws, accepting
\begin{lstlisting}
summon[((A | B) & C) =:= ((A & C) | (B & C))]  
summon[((A & B) | C) =:= ((A | C) & (B | C))]
\end{lstlisting}

We next consider performance consequences of distributivity in a type checker such as one in Scala (or, as we will see, TypeScript).
Consider the families of types $S_n$ and $T_n$ indexed on even numbers defined as
\begin{lstlisting}
S$_2$ = X$_1$ | X$_2$                          T$_2$ = X$_2$ | X$_1$
S$_{n+2}$ = S$_n$ & (X$_{2n-1}$ | X$_{2n}$)                    T$_{n+2}$ = T$_n$ & (X$_{2n}$ | X$_{2n-1}$)

\end{lstlisting}
That is, $S_n$ and $T_{n-1}$ only differ in the permutation of the disjuncts. Interestingly, the Scala compiler fails to prove \lstinline[language=scala]|S$_n$ =:= T$_n$| for as small as $n=6$. Moreover, the compiler runtime increases exponentially despite the incompleteness, with typecheking time for $n=26,28,30,32, 34$ being $13, 41, 135, 280, 640$ seconds. On the other hand, a verified orthologic algorithm \cite{guilloudVerifiedOptimizedImplementation2025} solves such subtyping querries in milliseconds, exhibiting quadratic time.

Currently, Scala type checker also accepts the following ``constructor conjunctivity'' law (generalizing the law~\eqref{eq:syntacticFun} for functions that we discussed in the introduction): if \lstinline|T[+X]| is a covariant type constructor, then \lstinline[language=scala]|T[A] & T[B] =:= T[A & B]|. Alarmingly, we discovered recently that this rule is unsound, leading to runtime error for programs that type check, such as this one:
%(\autoref{apdx:snippets}, \autoref{fig:scalaunsoundness}). 
\begin{lstlisting}[language=scala]
    trait L[+A]{val a:A}
    trait R[+B]{val b: B}
    class LR(val a: Int, val b: String) extends L[Int] with R[String]
    type E[+A] = L[A] | R[A]
    
    val x: E[Int] & E[String] = LR(4, "hi")
    val y: E[Int&String] = x
    val z: Int&String = y match
        case l : L[Int&String] => l.a
        case r : R[Int&String] => r.b
    z.toUpperCase
\end{lstlisting}
Further study is necessary to establish when such laws are sound. In this paper, we do not consider such problematic constructor conjunctivity laws. Our constructors are simply monotonic (covariant), antimonotonic (contravariant), or uninterpreted (invariant) in their arguments. The algorithms we present support efficient and complete reasoning about subtyping (and equivalence) and normalization for this model.

\paragraph{Typescript}

Typescript adds static types and typechecking to JavaScript. It supports union and intersection types, which form a distributive lattice. Analogously to Scala, TypeScript validates the following assertions: 
\begin{lstlisting}[language=pseudo]
type Eq<T, U> = T|U extends T&U? true : false;
type Assert<T extends true> = T;

type V1 = Assert<Eq<A|B, B|A>>;           type V1_ = Assert<Eq<A&B, B&A>>;
type V2 = Assert<Eq<(A|B)|C, A|(B|C)>>;   type V2_ = Assert<Eq<(A&B)&C, A&(B&C)>>;
type V3 = Assert<Eq<A|A, A>>;             type V3_ = Assert<Eq<A&A, A>>;
type V4 = Assert<Eq<A|(A&B), A>>;         type V4_ = Assert<Eq<A&(A|B), A>>;
type V5 = Assert<Eq<A|unknown, unknown>>; type V5_ = Assert<Eq<A&never, never>>;
type V6 = Assert<Eq<A|never, A>>;         type V6_ = Assert<Eq<A&unknown, A>>;

type Distributivity = Assert<Eq<(A&B)|C, (A|C)&(B|C)>>
type Distributivity_ = Assert<Eq<(A&B)|C, (A|C)&(B|C)>>
\end{lstlisting}
On the families $\Phi_n$ and $\Phi'_n$ we introduced in the previous section, Typescripts demonstrates significantly faster performances than Scala type checker, but still exhibits exponential behaviour, with type checking times for $n=26, 28, 30, 32$ being $1.5, 2.1, 3.1, 5.5$, and interrupts compilation with an error TS2590 for $n\geq 34$, when the size of types after normalization exceeds a certain limit.

On the other hand, Typescript does not support polymorphic function with lower bounds as in the above Scala example, nor nominal subtyping declarations.

\paragraph{Flow}
Flow~\cite{flow} is a type system for JavaScript based on flow typing. It also uses types that form a lattice, but it is \textit{not} distributive. Moreover, the subtype checking algorithm is highly incomplete: it does not even accept $\Phi_4 = \Phi'_4$ without user-written assertions, that is, it fails to prove
\begin{lstlisting}
X1|X2 & X3|X4 = X2|X1 & X4|X3
\end{lstlisting}
Note the difference of the situation compared to Scala and TypeScript, which implement distributivity and may compute a disjunctive normal form, which gives them a reason to be slow. Here, the type lattice of Flow is not claimed to be distributive, so there is no reason to expect slow performance: an approach based on lattices or ortholattices would yield a complete and fast subtyping algorithm.

Other languages have lattice-like subtyping relation. For example, Julia, Typed Racket, and Crystal offer union types to the user, but not intersection types. Kotlin uses intersection types internally, but does not expose them to the user. PHP offers union and intersection types, but they cannot be combined. In general, we believe that algorithms based on lattices with constructors would benefit implementations in these systems.

In summary, Scala and TypeScript implement distributivity law, but face slowdown in type checking, whereas Flow does not implement distributivity. We therefore feel justified in not assuming the general distributivity law.

\paragraph{Type Simplification}

Simplifying the representation of types can improve performances and give better messages to the user. All the above languages perform some kind of type normalization.
For example, it is natural to infer the type \lstinline{A|B} for an expression \lstinline{if e then a else b} where \lstinline{a:A} and \lstinline{b:B}. However, it will often be the case that \lstinline{A} and \lstinline{B} are the same type, for example \lstinline{Int}. In this case it is desirable to simplify \lstinline{Int|Int} to \lstinline{Int} before continuing to typecheck the program. 

Similarly, computing a normal form for types (that is, mapping equivalent types to the same representation) simplify algorithms and allow for various optimizations, such as the use of caching. However, simplification and normalization are not necessarily compatible: In Boolean algebra for example, CNF and DNF are normal forms but they are not simplifications, as they can increase (exponentially) the representation of an expression. Hence, languages such as TypeScript do not compute a full normal form and instead apply practical heuristics.

On the other hand, orthologic with monotonic functions admits an efficient and easy to implement normalization procedure that can never increase the size of types (presented in \autoref{sec:normalization}). Such normalization can serve as a basis for simplification of types in type checkers.

\subsection{Negation Types}
Negation types, which we denote \lstinline|~A| for a type \lstinline|A|, have been treated in type system designs 
\cite{castagnaProgrammingUnionIntersection2024, lionelparreauxHkusttacoMlscript2025, parreauxMLstructPrincipalType2022}, but have not been as widely adopted as union and intersection types. 
One may think that negation types typically bear little computational information that could ensure safety: What can be done on a value that is known to be ``anything but a string" that could not be done on a string? Hence, the utility of negation types resides more on the value of types as contract between the an implementation and a user. Nonetheless, there are some specific types whose negation do bear computational value, such as 
\begin{itemize}
    \item \lstinline|~Null|, the type of non-null values
    \item \lstinline|~Ref|, the type of values that are not references
    \item \lstinline|~(Any => Nothing)|, the type of values that are not functions
    \item \lstinline|~{bar:Any}|, the type of records that do not define the \lstinline|bar| field, allowing it to be safely introduced
\end{itemize}

TypeScript and Flow do support conditional types, which can be used to represent a form of type negation
\begin{lstlisting}
    function foo<A>(x: A extends number?never:A) : A {return x}
    foo("hello world") //compiles
    foo(4) //does not compile
    foo<unknown>(4) //compiles
\end{lstlisting}
As the last line shows, the behaviour relies on incomplete type inference and is not a true negation type. Negation types have been a long-time requested feature for TypeScript \cite{negatedTypesTS}.

\paragraph{Possible Use Case: Flow Types in Pattern Matching}
Consider the following program snippet:
\begin{lstlisting}
trait A
class B1 extends A
class ... extends A

def handleB1(x: A) =
  x match
    case _:B1 => "good"
    case _ => y
\end{lstlisting}
Currently, Scala infers {\tt String|A} as the result type of {\tt handleB1}, even though its semantics of pattern matching allows overlapping patterns and prescribes a top-down execution of cases. A system based on flow typing could thus apply the fact that when {\tt y} is returned, it does not have type {\tt B1}, which suggests that a more precise return type for {\tt handleB1} is \lstinline$String | (A & ~B)$. 

MLscript \cite{lionelparreauxHkusttacoMlscript2025} and its core subset MLstruct \cite{parreauxMLstructPrincipalType2022} offer negation types as well as union and intersections to assign more general types to expressions in a way similar to the above example, as witnessed by the following illustration:
\begin{lstlisting}
def flatMap2 f opt = 
  case opt of 
    Some -> f opt.value , 
    _    -> opt
flatMap2 : $\forall$ A, B. (A -> B) -> (Some[A] | (~Some[Any] & B)) -> B
\end{lstlisting}
In this example, the negation type allows to exclude from \lstinline|B| types for which \lstinline|opt| would be caught by the first case of the match pattern, but with a different type parameter.

CDuce \cite{castagnaProgrammingUnionIntersection2024, cduceCDuceCompiler2021} also uses negation types to assign more precise types to expressions. A simple example is the \lstinline|not| function:
\begin{lstlisting}
let not = fun x -> if x then false else true
\end{lstlisting}
If \lstinline|Falsy| is the type of values considered as wrong\footnote{For example in Javascript \lstinline|false|, \lstinline|""|, \lstinline|0|, \lstinline|-0|, \lstinline|0n|, \lstinline|undefined|, \lstinline|null|, and \lstinline|NaN|} it can be assigned the type 
\begin{lstlisting}
~Falsy -> false & Falsy -> true
\end{lstlisting}
which is more precise than \lstinline|bool -> bool|. This is also an example where distribution of intersection over function types (\autoref{eq:syntacticFun}) does not hold in the type system.

\paragraph{Possible Use Case: Unambiguous Overload}
Function overload allows to define different implementations of the same function based on the type of the value it is applied to. However, determining which implementation should be invoked can be ambiguous. Suppose \lstinline|AB <: A| and \lstinline|AB <: B|, and two function definitions
\begin{lstlisting}
def f(x: A)
def f(x: B)
\end{lstlisting}

Now if \lstinline|ab: AB|, resolving which \lstinline|f| should \lstinline|f(ab)| use is ambiguous. Such cases are typically rejected by compilers, or be otherwise resolved by a fixed notion priority relying on precision or scope. Negation types allow to give arbitrary priority for such cases, by defining instead (if the first implementation should have priority)
\begin{lstlisting}
def f(x: A)
def f(x: B & ~A)
\end{lstlisting}
Such use of negation thus enables expressive compile-time disambiguation of calls depending on the arguments they handle.

In general, negation provides a rich language to express disjointness of combinations of types. We thus also expect it to be of interest for match types in Scala \cite{DBLP:journals/pacmpl/BlanvillainBKO22}, which impose disjointness conditions on cases when matching on the type.

%Match types in Scala?

%sounds more like related work

%\paragraph{CDuce}

\subsection{Type Constructors and Typing Constraints}

Having motivated the benefits of using orthologic for type system design, we now turn to the possibilities that open up once we extend such efficient reasoning to support covariant, contravariant and invariant type constructors.

\paragraph{Record Types}
Records are ubiquitous in programming languages. We can naturally represent record types by combining intersection types and covariant type constructors as follows.
For every possible field $s$, define a type constructor \lstinline|R$_s$[+T]| with one covariant argument. Then, multi-fields record types are represented as the intersection of the record types of each field, so for example \lstinline|R$_\texttt{foo}$[String] & R$_\texttt{bar}$[Int]| represents the record type \lstinline|{foo: String, bar: Int}|. Given such definition, the usual depth, width, and permutation subtyping relations are provable from the laws of $\OLP$, so they are fully supported by our algorithms.

\paragraph{Nominal Subtyping and Traits}
Constraints naturally allow our system to express relations between user-defined types using classes, interfaces and other similar constructs. Consider the following example.
\begin{lstlisting}
trait S
trait T[A]
class U extends S, T[S]
\end{lstlisting}
We introduce three types \lstinline|S|, \lstinline|T[A]| and \lstinline|U| such that \lstinline|U <: S| and \lstinline|U <: T[S]|. These two conditions (or equivalently \lstinline|U <: S & T[S]|) can be added to the set of constraints in the program.
The situation is more complex if $U$ is a polymorphic type, as in the following  example:
\begin{lstlisting}
trait S[A]
trait T[A]
class U[A] extends S[A], T[S[A]]
\end{lstlisting}
The corresponding constraint becomes
$$
U(A) <: S(A) \land T(S(A))
$$
which should be understood as universally quantified over all $A$. Assumptions of this kind can be handled with algebraic subtyping, similarly as in \cite{dolanAlgebraicSubtypingDistinguished2017}, reducing general axioms into a form of a definition. For the above example, introduce an unconstrained symbol $U'(A)$, and define $U$ as $U(A) := U'(A) \land S(A) \land T(S(A))$, and desired property then follows by substitution, without the need for axioms. (The substitution is efficient in our algorithms because they support structure sharing.) For a formal statement and proof of this, see \autoref{apdx:proofs}, \autoref{thm:abstracttypesdefinitions}.
Note that the dual construction allows to define types with upper bounds.

\paragraph{Nominal Equirecursive Types}
Equirecursive types are characterized by a recursive relations. As an exmple, consider the recursively-defined type of JSON data in a Scala-like syntax:
\begin{lstlisting}
type JSON = Double | String | Boolean | Null | Map[String, JSON] | Seq[JSON]
\end{lstlisting}
Such recursive definitions using untagged unions are considered difficult to reason about, in contrast to their version that use tagged union (that said, TypeScript supports some non-nominal equirecursive types since TypeScript 3.7).
In our approach, it is straightforward to model such definitions without relying on unfolding: they become simple constraints. Under such semantics, the types are nominal in the sense that the above \lstinline|JSON| would not be considered equal to another type defined using an alpha-equivalent definition.

We also acknowledge that our approach does not generalize to automatic support for recursive definitions with type parameters. (Indeed, such definitions would correspond to universally quantified assumptions; it is easy to see that allowing universally quantified assumptions with free symbols yields full equational logic, which is undecidable.) A simple possibility would be for the programmer to manually instantiate type parameters for the cases of interest. 

\paragraph{Constrained Polymorphism}
Polymorphic functions need no introduction. Most language with subtyping offer \textit{bounded} polymorphism~\cite{daySubtypesVsWhere1995}, for example \lstinline|def foo[T<:B]|, where \lstinline|foo| is only polymorphic over subtypes of \lstinline|B|. Some languages also offer lower bounds. In our approach, this can be generalized to arbitrary constraints, which could be written, for example, as
\lstinline|def bar[T where F[T] <: G[T]]|, for arbitrary expressions \lstinline|F[T]| and \lstinline|G[T]|. Inside the body of the function, this amounts to typechecking with the additional assumption given by the constraint. At a call site, a subtyping check needs to be performed for the constraint, instantiated with the adequate type (where, in the simple case, the instantiation could be provided manually). Such \textit{constrained polymorphism} offers a lot of expressibility. Here are some examples:
\begin{lstlisting}
def f[S, T where S & T <: Nothing]  //only accepts pairs of disjoint types
def g[T where T <: Comparable[T]]  //F-bounded polymorphism is a special case
def h[T where T <: Comparable[T], Set[T] <: Comparable[Set[T]]]   
                                    //Multiple constraints are allowed
class List[A]:
   def toMap[S, T where A <: (S, T)] //Constraints parameters from outer scope
\end{lstlisting}
Scala allows to express similar constraints using its implicit resolution system with subtyping evidence, which comes at a cost in reliability and performances \cite{krikavaScalaImplicitsAre2019}.
%In contrast, these features come for free in a type system that is naturally able to reason with constraints.

Reasoning about subtyping in the presence of constraints has been studied in multiple occasions (see e.g \cite{oderskyTypeInferenceConstrained1999, remyAdvancedTopicsTypes2004, vytiniotisOutsideInXModularType2011}). Even if they are not offered to the user, they can also arise in internal representations, for example due to generalizaed algebraic datatypes (GADTs).

These examples give us hope that ground reasoning with type constructors as supported by our system can serve as a key part of an expressive type system implementation.

\section{Preliminaries on Ortholattices}
\label{sec:preliminaries}

We briefly review some necessary concepts of universal algebra. For a comprehensive introduction, we refer the reader to e.g. \cite[Chapter 2]{DBLP:books/daglib/0067494}.
\begin{definition}[Algebraic Signature]
    
An \textit{algebraic signature}, or simply \textit{signature} is a set of function symbols each annotated with an arity. For example, $+^2, \times^2, -^1, 1^0, 0^0$ is the signature of rings and fields.
\end{definition}

\begin{definition}[Lattices and Ortholattices]

A \textit{lattice} is an algebraic structure with base set $A$ and signature $\land^2, \lor^2$ satisfying laws V1-V4 and V1'-V4' of~\autoref{tab:algebraiclaws}.
Every lattice is also a partially ordered set (poset for short) under the relation $\leq$ such that
$$x \leq y \iff x = x \land y$$
(which is also equivalent to $y = y \lor y$ in the presence of lattice axioms). $x \land y$ is the greatest lower bound of $x$ and $y$, while $x \lor y$ is the least upper bound. A poset in which any pair of element admit a greatest lower bound and least upper bound is always a lattice. Types with the subtyping relationship are often designed to form a lattice.
A \textit{bounded lattice} is an algebraic structure $A, \land^2, \lor^2, \bot^0, \top^0$ which satisfies laws V1-V6 and V1'-V6' of \autoref{tab:algebraiclaws}. $\bot$ is the smallest element of the lattice, and $\top$ the largest.
An \textit{ortholattice} is an algebraic structure $A, \land^2, \lor^2, \neg^1, \top^0, \bot^0$ which satisfies the laws V1-V9 and V1'-V9' of \autoref{tab:algebraiclaws}.
\end{definition}

\begin{definition}[Monotonic Functions]
An {\LP} (resp. \BLP, \OLP) is a lattice (resp. bounded lattice, ortholattice) with countably many additional function symbols $f^{i,j,k}$, (for $i,j,k \in \mathbb{N}$ and $f$ in some infinite alphabet), each of arity $i+j+k$, satisfying for each such symbol the law V10 of \autoref{tab:monotonicitylaw}.
We say that $f^{i,j,k}$ is \textit{invariant} in the first $i$ arguments, \textit{covariant} (or monotonic) in the next $j$ arguments, and \textit{contravariant} (or antimonotonic) in the last $k$ arguments.
\end{definition}

\begin{table*}[bth]    
    \centering
    \begin{tabular}{r c}
         V10: & $ \forall n \in \lbrace 0,...,j\rbrace, y_n \leq y'_n \implies $  \\
              & $ \forall n \in \lbrace 0,...,k\rbrace, z'_n \leq z_n \implies$  \\
              & $ f^{i,j,k}(x_1, ..., y_1, ..., z_1, ...)  \leq 
                  f^{i,j,k}(x_1, ..., y'_1, ..., z'_1, ...)$  \\[3ex]

        V10': & $f^{i,j,k}(x_1, ..., y_1, ..., z_1, ...)  \leq 
                 f^{i,j,k}(x_1, ..., y_1 \lor y'_1, ..., z_1 \land z'_1, ...)$\\[3ex]
              
    \end{tabular}
    \vspace{2ex}
    \caption{Two equivalent characterization of the monotonicity law (\autoref{thm:variety}).}
    \label{tab:monotonicitylaw}
\end{table*}

\begin{definition}[Terms]
Let $X$ be an infinite, countable set of variables. For an algebra signature $S$, let $\T_S(X)$ be the set of terms generated by $X$ and $S$. Terms are trees whose leaves are labelled by members of $X$ and constant symbols of $S$ and nodes are labelled with function symbols of $S$ of arity $>0$.
For a class of algebras $K$, let $\simw{K}$ be the congruence relation on $\term{X}{K}$ defined as $s \simw{K} t \iff K \models s = v$ that is, $s = t$ holds in every member of $K$ under any interpretation of variable symbols. Note that for classes defined by equations (which is the case of all classes we consider), $s \simw{K} t$ if and only if $s = t$ is provable (in first order logic) from the axioms of $K$. For example, $\simw{\OLP}$ is the relation on $\term{X}{\OLP}$ such that $t_1 \simw{\OLP} t_2$ if and only if $t_1 = t_2$ it is provable from the axioms V1-V9, V1'-V9', V10.

For any class of algebras $K$ that is a subclass of the class of lattices (which will be the case of all classes considered in the present paper), we define similarly the relation $\leq_K$ on $\term{X}{K}$ as $s \simw{K} t \iff K \models s \leq v$. Note that $s \leq_K t \iff s \simw{K} s \land t \iff t \simw{K} s \lor t$.
\end{definition}

%The terminology for e can be conflicting. 
In universal algebra, elements of $\term{X}{\OLP}$, the set of trees over the language of an algebra, are often called \textit{terms}. %In a logical setting (such as in \cite{guilloudFormulaNormalizationsVerification2023, guilloudOrthologicAxioms2024}, they would be called \textit{formulas}. 
We follow this terminology. (The use of word \emph{term} is not to be confused with potential inhabitants of a type in a programming language; in this paper we do not formally study the relation $p:\tau$ between programs $p$ and the types $\tau$.)
We use $S, T, U$ to denote arbitrary terms. 
$A, B, C \in X$ denote variables and $F, G, H$ denote function symbols of $\OLP$.
%Note that two terms $s$ and $t$ are in the same equivalence class in $\F_K$ if and only if they are provably equal under the identities defining $K$. 

\begin{definition}[Word Problem]
The \textit{word problem} for a fixed class of algebra $K$, consists in deciding, for any pair $s, t \in \term{X}{K}$, if $s \simw{K} t$. 
A (finite) presentation for $K$ is a (finite) set of ground identities.
The \textit{entailment problem}\footnote{Also called uniform generalized ford problem in the field of universal algebra} consists in deciding, for any given finite presentation $A = \lbrace (s_1, t_1), ..., (s_n, t_n) \rbrace$ and pair of terms $s$ and $t$, if $s \simw{K|A} t$, where $K|A$ is the restriction of $K$ to algebras satisfying $s_1 = t_1, ..., s_n = t_n$.
\end{definition}

Note that for subclasses of the classes of lattices, we can equivalently phrase the problem as deciding if $s \leq_K t$, and relax presentations to contain inequalities rather than equalities.

\section{Entailment Problem for Ortholattice Types with Constructors}
\label{sec:generalizedwordproblem}
In this section we present a proof system for the entailment problem in \OLP, that is, the problem of deciding if a term $S$ is a subtype of another term $T$ under given constraints. Our development has been formalized in the Rocq proof assistant (see Supplementary material). In the text below, we annotate selected theorems and definitions with their corresponding identifiers in the formal development. For example\coqref{SCPlus\_to\_CFPlus} annotates the cut elimination theorem~\ref{thm:cutelim}.

\begin{theorem}[Polynomial algorithm for the entailment problem in \OLP]
There is an algorithm such that given
    \begin{itemize}
        \item $A$ a finite set of inequalities $(a_1 \leq b_1), ..., (a_m, b_m)$, where $a_i, b_i \in \term{X}{\OLP}$
        \item $S, T \in \term{X}{\OLP}$
    \end{itemize}
    decides if $w \leq_{\OLP|A} w'$ holds in time $\mathcal{O}(n^2(1+|A|)$, where $n$ is the total size of formulas in $A$, $S$ and $T$.
\end{theorem}

Throughout this section, fix the set of axioms $A$. To ease notation, we refine $\OLP$ to stand for the variety of all ortholattices with monotonic functions \textit{satisfying axioms in $A$}, any further definition and theorem in this section can be implicitly understood as parametric on $A$.

To prove the above theorem, we first construct a proof system for pairs of $\OLP$ terms such that a pair of terms $(s, t)$ is provable if and only if $s \leq_\OLP t$. The system builds on the previously introduced one \cite{guilloudOrthologicAxioms2024}, with the key extension being monotonic functions. This system contains a rule (Cut), corresponding to transitivity and making proof search difficult. The core of the proof consists in showing that this rule can be eliminated. Finally, the resulting proof system admits a \textit{subformula property}, allowing us to build an efficient proof search algorithm.

\begin{definition}[Annotated Terms, Sequents]
    If $T$ is a term, $T^L$ and $T^R$ are \textit{annotated terms}. Symbols $\Gamma, \Delta$  range over annotated terms. A sequent is an unordered pair (multiset of size two) of annotated terms $(\Gamma, \Delta)$, often written without parentheses.
\end{definition}

\begin{definition}[\SCP\coqref{SCPlus}]
    The proof system $\SCP$ is presented in \autoref{fig:proofSystem}. Define the relation 
    $$\vdash_\SCP \Gamma, \Delta$$
    to hold if and only if there exists a proof of $\Gamma, \Delta$ in $\SCP$ using axioms in $A$.
    
\end{definition}

\begin{figure}[ht]
    \begin{tabular}{c c}
    \ \\
        \multicolumn{2}{c}{            
            \AxiomC{}
            \RightLabel{\scrule{ Hyp}}
            \UnaryInfC{$S^L, S^R$}
            \DisplayProof
        }\\[3ex]

        \multicolumn{2}{c}{
            \AxiomC{$\Gamma, T^R$}
            \AxiomC{$T^L, \Delta$}
            \RightLabel{\scrule{ Cut}}
            \BinaryInfC{$\Gamma, \Delta$}
            \DisplayProof
        }\\[3ex]
    
        \multicolumn{2}{c}{
            \AxiomC{$\Gamma, \Gamma$}
            \RightLabel{\scrule{ Replace}}
            \UnaryInfC{$\Gamma, \Delta$}
            \DisplayProof
        }\\[3ex]
        
        \AxiomC{}
        \RightLabel{\scrule{ LeftBot}}
        \UnaryInfC{$\bot^L, \Delta$}
        \DisplayProof &
        \AxiomC{}
        \RightLabel{\scrule{ RightTop}}
        \UnaryInfC{$\Gamma, \top^R$}
        \DisplayProof
        \\[3ex]

        \AxiomC{$\Gamma, S^L$}
        \RightLabel{\scrule{ LeftAnd}}
        \UnaryInfC{$\Gamma, (S \land T)^L$}
        \DisplayProof &
        \AxiomC{$\Gamma, S^R$}
        \AxiomC{$\Gamma, T^R$}
        \RightLabel{\scrule{ RightAnd}}
        \BinaryInfC{$\Gamma, (S \land T)^R$}
        \DisplayProof
        \\[3ex]

        \AxiomC{$\Gamma, S^L$}
        \AxiomC{$\Gamma, T^L$}
        \RightLabel{\scrule{ LeftOr}}
        \BinaryInfC{$\Gamma, (S \lor T)^L$}
        \DisplayProof &
        \AxiomC{$\Gamma, S^R$}
        \RightLabel{\scrule{ RightOr}}
        \UnaryInfC{$\Gamma, (S \lor T)^R$}
        \DisplayProof
        \\[3ex]

        \AxiomC{$\Gamma, S^R$}
        \RightLabel{\scrule{ LeftNot}}
        \UnaryInfC{$\Gamma, (\neg S)^L$}
        \DisplayProof &
        \AxiomC{$\Gamma, S^L$}
        \RightLabel{\scrule{ RightNot}}
        \UnaryInfC{$\Gamma, (\neg S)^R$}
        \DisplayProof \\[3ex]

        \multicolumn{2}{c}{
            \AxiomC{$S_{x_1}^L T_{x_1}^R$}
            \AxiomC{$T_{x_1}^L S_{x_1}^R$\quad ...}
            \AxiomC{$S_{y_1}^L T_{y_1}^R$\quad ...}
            \AxiomC{$T_{z_1}^L S_{z_1}^R$\quad ...}
            \RightLabel{\textup{  \scrule{$F$-rule} }}
            \QuaternaryInfC{$F(S_{x_1}, ..., S_{y_1}, ..., S_{z_1}, ...)^L, F(T_{x_1}, ..., T_{y_1}, ..., T_{z_1}, ...)^R$}
            \DisplayProof 
        }\\[3ex]
        
        \multicolumn{2}{c}{            
            \AxiomC{\phantom{$\Gamma$}}
            \RightLabel{\scrule{ \scrule{Axiom} \quad if $(S, T) \in A$}}
            \UnaryInfC{$S^L, T^R$}
            \DisplayProof
        }
\ \\

    \end{tabular}
    \caption{Deduction rules of Orthologic. Each holds for arbitrary $\Gamma$, $\Delta$, $S$, $T$}    

\label{fig:proofSystem}
\end{figure}

The proof system is inspired by \cite{guilloudOrthologicAxioms2024}, with two key difference. In the system from \cite{guilloudOrthologicAxioms2024}, sequents have \textit{at most two} formulas rather than exactly two, with an implicit contraction rule $\frac{\Gamma}{\Gamma, \Gamma}$ that lead to some edge case not being considered. This was later fixed in the formalization of \cite{guilloudVerifiedOptimizedImplementation2025} by making the rule explicit, and with a third kind of annotated term $N$ denoting the absence of term. Our version is simpler, replacing the Weaken and Contract rules from \cite{guilloudVerifiedOptimizedImplementation2025} by the single \scrule{Replace} rule, and only considers sequents with always exactly two annotated terms, without $N$.

Crucially, our system also includes the  \scrule{$F$-rule}, expressing the monotonicity of functions symbols 
$$F^{i,j,k}(x_1, ..., x_i, y_1, ..., y_j, z_1, ..., z_k)$$
that is invariant in the first $i$ argument, covariant in the next $j$ arguments and contravariant in the last $k$ arguments. To limit notational overload, we omit the indices $i,j,k$ entirely.

\begin{theorem}[Soundness\coqref{SCPlus\_soundness}]
\label{thm:soundness}
    Let $S, T \in \term{X}{\OLP}$ be terms. If $\vdash_\SCP S^L, T^R$ then $S \leq T$ in every \OLP.
\end{theorem}
\begin{proof}
    By induction on the proof of $S^L, T^R$, strengthening the induction hypothesis so that 
    $$\vdash_\SCP S^L, T^L \implies S \leq \neg T $$
    and similarly $\vdash_\SCP S^R, T^R \implies \neg S \leq T $ (in all ortholattices).
\end{proof}

\begin{theorem}[Completeness\coqref{SCPlus\_completeness}]
\label{thm:completeness}
    Let $S, T \in \term{X}{\OLP}$ be terms. If $S \leq T$ in every $\OLP$, then $\vdash_\SCP S^L, T^R$.
\end{theorem}
\begin{proof}
    For $S, T \in \term{X}{\OLP}$, let $S \dashvdash T$ be the relation defined by 
    $$S \dashvdash T \iff (\vdash_\SCP S^L, T^R)\, \&\, (\vdash_\SCP T^L, S^R)$$
    It is not difficult to prove that $\dashvdash$ is a congruence relation, and that $\term{X}{\OLP}_\slash{\dashvdash}$ is itself an ortholattice.
    Hence, if $S \leq T$ holds in all ortholattices, it holds in particular in $\term{X}{\OLP}_\slash{\dashvdash}$, and hence by definition $\vdash_\SCP S^L, T^R$.
\end{proof}

Soundness and completeness give us an alternative characterization of the generalized word problem for $\OLP$: an inequality $S \leq_\OLP T$ is true in all ortholattices if and only if it is provable in $\SCP$. 

Towards a proof search procedure for $\SCP$, note that all rules except for the \scrule{Cut} rule have the subterm property: all terms appearing in the premises of a rule are subterms of the terms appearing in the conclusion. We aim to use this property to guide proof search, but we first need to show that the \scrule{Cut} rule can be suitably restricted.

\begin{definition}[\CFP\coqref{CFPlus}]
    The proof system \CFP contains the rules of \SCP with the exception \scrule{Axiom} and \scrule{Cut},  and adds instead the rule
    \begin{center}
        \AxiomC{$\Gamma, S^R$}
        \AxiomC{$T^L, \Delta$}
        \RightLabel{\scrule{AxiomCut \quad (with $(S, T) \in A$)}}
        \BinaryInfC{$\Gamma, \Delta$}
        \DisplayProof
    \end{center}
    
\end{definition}

\begin{theorem}[Partial \scrule{Cut} Elimination\coqref{SCPlus\_to\_CFPlus}] 
\label{thm:cutelim}
    A sequent has a proof in $\SCP$, if and only if it has a proof in \CFP.
\end{theorem}

\begin{proof}
For the `if'  direction, it suffices to note that AxiomCut is indeed a restriction of the \scrule{Axiom} and \scrule{Cut} rule.
For the `only if' direction, first note that the original form of the \scrule{Axiom} rule can be easily recovered with \scrule{Hyp}, as
    \begin{center}
        \AxiomC{}
        \RightLabel{\scrule{Axiom}}
        \UnaryInfC{$S^L, T^R$}
        \DisplayProof
    \end{center}
is equivalent to
    \begin{center}
        \AxiomC{}
        \RightLabel{\scrule{Hyp}}
        \UnaryInfC{$S^L, S^R$}
        \AxiomC{}
        \RightLabel{\scrule{Hyp}}
        \UnaryInfC{$T^L, T^R$}
        \RightLabel{\scrule{AxiomCut \quad (with $(S, T) \in A$).}}
        \BinaryInfC{$S^L, T^R$}
        \DisplayProof
    \end{center}

The \scrule{Cut} elimination procedure extends the inductive argument of \cite{guilloudOrthologicAxioms2024}. 
In particular, consider the topmost occurence of the $Cut$ rule in the proof of a sequent $\Gamma, \Delta$. 
\begin{center}
    \AxiomC{$\A$}
    \UnaryInfC{$\Gamma, T^R$}
    \AxiomC{$\B$}
    \UnaryInfC{$T^L, \Delta$}
    \RightLabel{\scrule{ Cut}}
    \BinaryInfC{$\Gamma, \Delta$}
    \DisplayProof
\end{center}
We need to recursively eliminate the \scrule{Cut} rule from such proofs, where by induction $\A$ and $\B$ are proofs in \CFP. We proceed by case analysis on $\A$ and $\B$. For example,
\begin{center}
\begin{tabular}{c c c}
    \AxiomC{$\A'$}
    \UnaryInfC{$S^L, T^R$}
    \RightLabel{\scrule{RightNot}}
    \UnaryInfC{$\neg S^R, T^R$}
    \AxiomC{$\B'$}
    \UnaryInfC{$T^L, \Delta$}
    \RightLabel{\scrule{Cut}}
    \BinaryInfC{$\neg S^R, \Delta$}
    \DisplayProof 
    
    &
    
    \hspace{0.5em} $\hookrightarrow$ \hspace{0.5em}

    &
    
    \AxiomC{$\A'$}
    \UnaryInfC{$S^L, T^R$}
    \AxiomC{$\B'$}
    \UnaryInfC{$T^L, \Delta$}
    \RightLabel{\scrule{Cut}}
    \BinaryInfC{$S^L, \Delta$}
    \RightLabel{\scrule{RightNot}}
    \UnaryInfC{$S^R, \Delta$}
    \DisplayProof 
    \\[2ex]
\end{tabular}
\end{center}
Is a valid transformation because the occurence of \scrule{Cut} in the resulting proof applies to smaller proofs, and hence by induction can be eliminated.

The proof contains many cases. We do not present all of them, redirecting the reader to our formalization in Rocq and
a related proof of \cite[Theorem 3.8]{guilloudOrthologicAxioms2024}. We next highlight the cases involving the new \scrule{$F$-rule}.

Without loss of generality, suppose that $\A$ is an instance of the  \scrule{$F$-rule} , that is
\begin{center}
    \AxiomC{$S_{x}^L T_{x}^R$}
    \AxiomC{$T_{x}^L S_{x}^R$}
    \AxiomC{$S_{y}^L T_{y}^R$}
    \AxiomC{$T_{z}^L S_{z}^R$}
    \RightLabel{\textup{  \scrule{$F$-rule} }}
    \QuaternaryInfC{$F(S_{x}, S_{y}, S_{z})^L, F(T_{x}, T_{y}, T_{z})^R$}
    \AxiomC{$\B$}
    \UnaryInfC{$F(T_{x}, T_{y}, T_{z})^L, \Delta$}
    \RightLabel{\scrule{ Cut}}
    \BinaryInfC{$F(S_{x}, S_{y}, S_{z})^L, \Delta$}
    \DisplayProof
\end{center}
We abbreviate $F(S_{x}, S_{y}, S_{z})$ as $F(\vec{S})$, and $F(T_{x}, T_{y}, T_{z})$ as  as $F(\vec{T})$.

\textbf{Case \scrule{Hyp}.} If $\B$ is an instance of the \scrule{Hyp} rule,
\begin{center}
\begin{tabular}{c c c}
    \AxiomC{$\A$}
    \UnaryInfC{$F(\vec{S})^L, F(\vec{T})^R$}
    \AxiomC{}
    \RightLabel{\scrule{ Hyp}}
    \UnaryInfC{$F(\vec{T})^L, F(\vec{T})^R$}
    \RightLabel{\scrule{ Cut}}
    \BinaryInfC{$F(\vec{S})^L, F(\vec{T})^R$}
    \DisplayProof &
    
    $\hookrightarrow$ &
    
    \AxiomC{$\A$}
    \UnaryInfC{$F(\vec{S})^L, F(\vec{T})^R$}
    \DisplayProof
    \\[2ex]
\end{tabular}
\end{center}

\textbf{Case \scrule{Replace}.} If $\B$ is an instance of the \scrule{Replace} rule, there are two cases, depending on whether the replaced formula is the \scrule{Cut} formula or not. First:
\begin{center}
\begin{tabular}{c c c}
    \AxiomC{$\A$}
    \UnaryInfC{$F(\vec{S})^L, F(\vec{T})^R$}
    \AxiomC{$\B'$}
    \UnaryInfC{$\Delta, \Delta$}
    \RightLabel{\scrule{ Replace}}
    \UnaryInfC{$F(\vec{T})^L, \Delta$}
    \RightLabel{\scrule{ Cut}}
    \BinaryInfC{$F(\vec{S})^L, \Delta$}
    \DisplayProof &
    
    $\hookrightarrow$ &
    
    \AxiomC{$\B'$}
    \UnaryInfC{$\Delta, \Delta$}
    \RightLabel{\scrule{ Replace}}
    \UnaryInfC{$F(\vec{S})^L, \Delta$}
    \DisplayProof
    \\[2ex]
\end{tabular}
\end{center}
Second:

\begin{center}
    \AxiomC{$\A$}
    \UnaryInfC{$F(\vec{S})^L, F(\vec{T})^R$}
    \AxiomC{$\B'$}
    \UnaryInfC{$F(\vec{T})^L, F(\vec{T})^L$}
    \RightLabel{\scrule{ Replace}}
    \UnaryInfC{$F(\vec{T})^L, \Delta$}
    \RightLabel{\scrule{ Cut}}
    \BinaryInfC{$F(\vec{S})^L, \Delta$}
    \DisplayProof
\end{center}
Now, there are only 2 rules that can conclude with $F(\vec{T})^L, F(\vec{T})^L$: 
Replace and CutAxiom. But two consecutive \scrule{Replace} rules can be replaced by a single one. Finally

\begin{center}
\begin{tabular}{c c c}
    \AxiomC{$\A$}
    \UnaryInfC{$F(\vec{S})^L, F(\vec{T})^R$}
    \AxiomC{$\B'$}
    \UnaryInfC{$F(\vec{T})^L, U^R$}
    \AxiomC{$\B''$}
    \UnaryInfC{$V^L, F(\vec{T})^L$}
    \RightLabel{\scrule{ CutAxiom$(U, V)$}}
    \BinaryInfC{$F(\vec{T})^L, F(\vec{T})^L$}
    \RightLabel{\scrule{ Replace}}
    \UnaryInfC{$F(\vec{T})^L, \Delta$}
    \RightLabel{\scrule{ Cut}}
    \BinaryInfC{$F(\vec{S})^L, \Delta$}
    \DisplayProof \\[2ex]
    
    $\hookrightarrow$ \\[2ex]
    
    \AxiomC{$\A$}
    \UnaryInfC{$F(\vec{S})^L, F(\vec{T})^R$}
    \AxiomC{$\B'$}
    \UnaryInfC{$F(\vec{T})^L, U^R$}
    \RightLabel{\scrule{ Cut}}
    \BinaryInfC{$F(\vec{S})^L, U^R$}
    
    \AxiomC{$\A$}
    \UnaryInfC{$F(\vec{S})^L, F(\vec{T})^R$}
    \AxiomC{$\B''$}
    \UnaryInfC{$F(\vec{T})^L, V^L$}
    \RightLabel{\scrule{ Cut}}
    \BinaryInfC{$V^L, F(\vec{S})^L$}
    
    \RightLabel{\scrule{ CutAxiom$(U, V)$}}
    \BinaryInfC{$F(\vec{S})^L, F(\vec{S})^L$}
    \RightLabel{\scrule{ Replace}}
    \UnaryInfC{$F(\vec{S})^L, \Delta$}
    \DisplayProof
    \\[2ex]
\end{tabular}
\end{center}
Provides the desired transformation.

\textbf{Case Left and Right rules.} Note that since the \scrule{Cut} formula is $F(\vec{T})$, if $\B$ ends with \scrule{LeftAnd}, \scrule{RightAnd}, \scrule{LeftOr}, \scrule{RightOr} \scrule{LeftNot} or \scrule{RightNot}, the principal formula cannot be the same as the \scrule{Cut} formula. Hence, the translation is independent of $\A$ and similar to the \scrule{RightNot} case shown above.

\textbf{Case  \scrule{$F$-rule} .} If $G$ is a function symbols of \OLP and $\B$ ends with an instance of the $G$-rule, for some function symbol $G$, then necessarily $F = G$. Hence:
\begin{center}
\resizebox{\textwidth}{!}{
\begin{tabular}{c c c}

    \AxiomC{$\A_1$}
    \UnaryInfC{$S_{x}^L, T_{x}^R$}    
    \AxiomC{$\A_2$}
    \UnaryInfC{$T_{x}^L, S_{x}^R$}
    \AxiomC{$\A_3$}
    \UnaryInfC{$S_{y}^L, T_{y}^R$}
    \AxiomC{$\A_4$}
    \UnaryInfC{$T_{z}^L, S_{z}^R$}
    \RightLabel{\scrule{ \scrule{$F$-rule} }}
    \QuaternaryInfC{$F(S_{x}, S_{y}, S_{z})^L, F(T_{x}, T_{y}, T_{z})^R$}

    \AxiomC{$\B_1$}
    \UnaryInfC{$T_{x}^L, \theta_{x}^R$}    
    \AxiomC{$\B_2$}
    \UnaryInfC{$\theta_{x}^L, T_{x}^R$}
    \AxiomC{$\B_3$}
    \UnaryInfC{$T_{y}^L, \theta_{y}^R$}
    \AxiomC{$\B_4$}
    \UnaryInfC{$\theta_{z}^L, T_{z}^R$}
    \RightLabel{\scrule{ \scrule{$F$-rule} }}
    \QuaternaryInfC{$F(T_{x}, T_{y}, T_{z})^L, F(\theta_{x}, \theta_{y}, \theta_{z})^R$}
    
    \RightLabel{\scrule{ Cut}}
    \BinaryInfC{$F(S_{x}, S_{y}, S_{z})^L, F(\theta_{x}, \theta_{y}, \theta_{z})^R$}
    \DisplayProof\\[6ex]
    
    $\hookrightarrow$ \\

    \AxiomC{$\A_1$}
    \UnaryInfC{$S_{x}^L, T_{x}^R$}   
    \AxiomC{$\B_1$}
    \UnaryInfC{$T_{x}^L, \theta_{x}^R$}   
    \RightLabel{\scrule{ Cut}}
    \BinaryInfC{$S_{x}^L, \theta_{x}^R$} 
    
    \AxiomC{$\B_2$}
    \UnaryInfC{$\theta_{x}^L, T_{x}^R$}
    \AxiomC{$\A_2$}
    \UnaryInfC{$T_{x}^L, S_{x}^R$}
    \RightLabel{\scrule{ Cut}}
    \BinaryInfC{$\theta_{x}^L, S_{x}^R$} 
    
    \AxiomC{$\A_3$}
    \UnaryInfC{$S_{y}^L, T_{y}^R$}
    \AxiomC{$\B_3$}
    \UnaryInfC{$T_{y}^L, \theta_{y}^R$}
    \RightLabel{\scrule{ Cut}}
    \BinaryInfC{$S_{y}^L, \theta_{y}^R$} 
    
    \AxiomC{$\B_4$}
    \UnaryInfC{$\theta_{z}^L, T_{z}^R$}
    \AxiomC{$\A_4$}
    \UnaryInfC{$T_{z}^L, S_{z}^R$}
    \RightLabel{\scrule{ Cut}}
    \BinaryInfC{$\theta_{z}^L, S_{z}^R$}

    \RightLabel{\scrule{ \scrule{$F$-rule} }}
    \QuaternaryInfC{$F(S_{x}, S_{y}, S_{z})^L, F(\theta_{x}, \theta_{y}, \theta_{z})^R$}
    
    \DisplayProof\\[4ex]
\end{tabular}
}
\end{center}
Finally, if $F$ is itself of the shape of \scrule{AxiomCut}:
\begin{center}
\begin{tabular}{c c c}
    \AxiomC{$\A$}
    \UnaryInfC{$F(\vec{S})^L, F(\vec{T})^R$}
    \AxiomC{$\B'$}
    \UnaryInfC{$F(\vec{T})^L, U^R$}
    \AxiomC{$\B''$}
    \UnaryInfC{$V^L, \Delta$}
    \RightLabel{\scrule{ CutAxiom$(U, V)$}}
    \BinaryInfC{$F(\vec{T})^L, \Delta$}
    \RightLabel{\scrule{ Cut}}
    \BinaryInfC{$F(\vec{S})^L, \Delta$}
    \DisplayProof \\[2ex]
    
    $\hookrightarrow$ \\[2ex]
    
    \AxiomC{$\A$}
    \UnaryInfC{$F(\vec{S})^L, F(\vec{T})^R$}
    \AxiomC{$\B'$}
    \UnaryInfC{$F(\vec{T})^L, U^R$}
    \RightLabel{\scrule{ Cut}}
    \BinaryInfC{$F(\vec{S})^L, U^R$}
    \AxiomC{$\B''$}
    \UnaryInfC{$V^L, \Delta$}
    \RightLabel{\scrule{ CutAxiom$(U, V)$}}
    \BinaryInfC{$F(\vec{S})^L, \Delta$}
    \DisplayProof \\[2ex]
\end{tabular}
\end{center}
Hence, the proof system of \OLP admits partial \scrule{Cut} elimination.
\end{proof}
The next observation is that the entire proof system admits the subterm property (including the terms in $A$).
\begin{corollary}
    If a sequent $\Gamma, \Delta$ has a proof using axioms from $A$, then it has a proof using only subterms of $\Gamma, \Delta$ and of axioms in $A$.
\end{corollary}
\begin{proof}
    By induction on the structure of the proof, noting that all rules, when looked from conclusion to premises, never introduce terms that are not subterm of the conclusion or in the axioms.
\end{proof}

\section{Subtyping Algorithm}
\label{sec:subalgorithm}
We are finally ready to express our algorithm for checking inequalities in $\OLP$. We expect that its application will lead to predictable subtyping algorithms.

\begin{theorem}
\label{thm:proofsearch}
    There is a proof search procedure for $\OLP$ running in time $\mathcal{O}(n^2|A|)$, where $|A|$ is the number of given axioms and $n$ the total size of the problem.
\end{theorem}

\begin{proof}
We proceed by reducing proof search to validity of a set of propositional Horn clauses. 
For a sequent $\Gamma, \Delta$, let $\mathcal S(\Gamma, \Delta)$ be the set of all sequents that can be built from subterms of $\Gamma, \Delta$ and $A$. Note that there are at most $\sizeOf{S}+\sizeOf{A}$ such subterms and twice as many annotated terms, and hence 
$$\sizeOf{\mathcal S(\Gamma, \Delta)} \leq 4(\sizeOf{S}+\sizeOf{A})^2$$ 

Now, consider $\mathcal S(\Gamma, \Delta)$ as a set of propositional variables representing whether the corresponding sequent has a proof. Then, observe that every instance of a deduction rule whose premises and conclusion are in $\mathcal S(\Gamma, \Delta)$ corresponds to a Horn clause with $0$, $1$ or $2$ antecedents. Let $\clausesS{\Gamma, \Delta}$ be the set of all such clauses.
It follows that a sequent $S$ has a proof if and only if its corresponding variable is a logical consequence of all the clauses in $\clausesS{\Gamma, \Delta}$.

To bound the size of $\clausesS{\Gamma, \Delta}$, we now wish to count, for an arbitrary sequent $s \in \mathcal S(\Gamma, \Delta)$, how many Horn clauses can have $s$ as their conclusion.
Observe that the parameters of every rule but the \scrule{AxiomCut} rule, are uniquely determined by its conclusion, and the \scrule{AxiomCut} can be applied in as many different ways as there are axioms. Hence, any sequent $s$ can only be the conclusion of \textit{at most} $9+\sizeOf{A}$ different clauses, where $9$ is the number of deduction rules in $\CFP$ other than \scrule{AxiomCut}, and it follows that $|\clausesS{\Gamma, \Delta}| \leq |\mathcal S(\Gamma, \Delta)|(9 + |A|)$. Using $n = \sizeOf{(\Gamma, \Delta)} + \sizeOf{A}$, we obtain
$$|\clausesS{\Gamma, \Delta}| = \mathcal O(n^2(1+|A|))$$
Note that each clause contains at most 3 literals (one conclusion and two antecedents).

In \autoref{alg:pseudocode} we present pseudocode for the computation of these clauses. Each case from line 14 to 33 corresponds to a specific deduction rule. By \cite{dowlingLineartimeAlgorithmsTesting1984}, entitlement in Horn clauses can be decided in linear time using unit propagation. Hence, we conclude that deciding if an $\OL$-sequent has a proof is decidable in time $\mathcal O(n^2(1+|A|))$. Note that SAT solving for Horn clause typically proceeds by constructing a dependency graph, and we could of course build and solve the graph directly instead of using Horn clauses in \autoref{alg:pseudocode}. We chose to present reduction to Horn clauses instead, as we find it more compact and insightful.

\end{proof}
\begin{algorithm}[hbt]
\DontPrintSemicolon
\caption{Constructing the Horn clauses}\label{alg:pseudocode}
    \textbf{type} Formula$^*$ \cmnt{an annotated formula or a special None value}
    \textbf{type} Sequent $\gets$ (Formula$^*$, Formula$^*$)\; 
    \textbf{type} Clause $\gets$ (Sequent, Set[Sequent]) \cmnt{a Horn clause}
    A: Set[Sequent] $\gets$ Input \cmnt{set of axioms}
    conjecture: Sequent $\gets$ Input \;
    AxFormulas: Set[Formula] $\gets \bigcup \{ \{ a, b \} \mid \{ a^\Box, b^\circ \} \in \mbox{A} \}$ \cmnt{formulas from axiom sequents} 
    clauses : Set[Clause] $\gets$ Set.empty\;
    visited: Set[Sequent] $\gets$ Set.empty\;
    \procedure{addClause(s: Sequent, premises: Set[Sequent]): Unit}{
        clauses += (s, premises)\;
        premises.foreach(s => visit(s))
    }
    \procedure{visit(s: Sequent): Unit}{
        \If{visited.contains(s)}{\Return ()}
        \lineElseIf{s $\in$ A}{clauses += s}
        \Else{
            visited.add(s)\;
            \Switch(\tcp*[f]{non-exclusive switch, execute all}){s}{
                \Case{(None, $\Delta$)}{addClause(s, Set(($\Delta$, $\Delta$)))\cmnt{Impicit contraction}}
                \Case{$\phi^L$, $\phi^R$}{addClause(s, Set())}
                \Case{($\Gamma$, $\Delta$)}{addClause(s, Set(($\Gamma$, None)))}
                \Case{($(\neg\phi)^L, \Delta$}{addClause(s, Set(($\phi^R$, $\Delta$)))}
                \Case{($(\phi\land\psi)^L, \Delta$}{
                    addClause(s, Set(($\phi^L$, $\Delta$)))\;
                    addClause(s, Set(($\psi^L$, $\Delta$)))
                }
                \Case{($(\phi\lor\psi)^L, \Delta$}{
                    addClause(s, Set(($\phi^L$, $\Delta$), ($\psi^L$, $\Delta$)))
                }
                ...  \cmnt{analogous Right cases for $\Gamma$}
                ...  \cmnt{analogous Left and Right cases for $\Delta$}
                \Case{($\Gamma$, $\Delta$)}{
                    AxFormulas.foreach((x:Formula) $\rightarrow$ \;
                        \quad addClause(s, Set(($\Gamma$, x$^R$), (x$^L$, $\Delta$)))\;
                        \quad addClause(s, Set(($\Delta$, x$^R$), (x$^L$, $\Gamma$)))\;
                    )
                } 
            }
        }
    }
    visit(conjecture)\;
    \Return clauses
\end{algorithm}

\subsection{Additional Comments on Algebraic aspects}
We briefly highlight the interpretation of the results above in terms of universal algebra.

A \textit{variety} is a class of algebras with signature $S$ defined by a set of universally quantified identities of the form $s = t$, where $s, t \in \term{X}{S}$.
Lattices, Bounded Lattices and Ortholattices are clearly varieties, but not \LP, \BLP and \OLP, since the axiom $V10$ is not an identity\footnote{It is a quasi-identity, which characterize quasivarieties}. The following theorem gives an alternative characterization of $V10$ that is an identity.

\begin{theorem}{\label{thm:variety}}
    \LP, \BLP and \OLP are varieties.
\end{theorem}
\begin{proof}
    For every symbol $f$, axiom V10 is provably equivalent to V10' under lattice laws:
    \begin{itemize}
        \item V10 $\implies$ V10'. in $V10$, instantiate $y'_j$ by $y_j \lor y'1$ and $z'_j$ by $z_k \land z'k$. Since $y_j \leq_L y_j \lor y'_j$ and $z'_j \leq_L z_k \land z'k$, the conditions of V10 are fulfilled and we conclude.

        \item V10' $\implies$ V10. Since by assumption $y_j \leq y'_j$, we have $y_j \lor y'_j = y'j$. Similarly since by assumption $z'_j \leq z_j$, we have $z_j \land z'_j = z'j$, so we conclude.
        
    \end{itemize}
    Hence the class \LP{} (resp. \BLP, \OLP) is characterized by the axioms of lattices (resp. bounded lattices, ortholattices) and V10'.
\end{proof}

The \textit{free algebra} $\F_V(X)$ of a variety $V$ is the algebra such that, for any $\A \in V$ and any map $f : X \to \A$, $f$ extends uniquely to a homomorphism $\F_V(X) \to \A$. It is a theorem that $\F_V(X)$ is isomorphic to $\term{X}{S}\slashsimw{V}$ whose members are the equivalence classes of $\term{X}{V}$ by $\simw{V}$.

\begin{lemma}[Soundness and Completeness of \CFP]
\label{lem:cfp_sound_complete}
Let $\dashvdash_\CFP$ be the relation such that $S \dashvdash_\CFP T$ if and only if both $\vdash_\CFP S^L, T^R$ and $\vdash_\CFP T^L, S^R$ hold.
    For all $S, T \in \term{X}{\OLP}$, 
    $$(S \simw{V} T) \iff (S \dashvdash_\CFP T)$$
\end{lemma}
\begin{proof}
    Follows from \autoref{thm:soundness}, \autoref{thm:completeness} and \autoref{thm:cutelim}.
\end{proof}

So $\F_\OLP(X)$ is isomorphic to $\term{X}{\OLP}_\slash{\dashvdash_\CFP}$. In essence, we have constructed a syntactic and efficiently computable representation of $\F_{\OLP}(X)$, where for reminder \OLP is the variety of ortholattices with monotonic function symbols satisfying an arbitrary finite set of axioms $A$.
Note also that all theorems of this section generalize straightforwardly to \LP and \BLP (with axioms).

\section{Normal Form Algorithm for Ortholattices with Functions}
\label{sec:normalization}

In the present section, we show that $\OLP$ admits an efficiently computable \textit{normal form function}, which can be used to normalize types. The developments in this section generalize those of \cite{freeseFreeLattices1995, guilloudFormulaNormalizationsVerification2023, brunsFreeOrtholattices1976}. Whereas not all results in this section were formalized in Rocq, the proof-producing approach of \cite{guilloudVerifiedOptimizedImplementation2025} can be used to efficiently check the correctness of computed normal forms using the formalized proof system for $\OLP$ we presented in \autoref{sec:generalizedwordproblem}.

\begin{definition}[Minimal Form]\label{def:minimalform} Let $\A$ be an algebra.
    A term $T \in \term{X}{\A}$ is in \textit{minimal form} if there exists no $T$' such that $\sizeOf{T'} < \sizeOf{T}$ with $T' \simw{\A} T$.
\end{definition}

\begin{definition}[Normal Form Function]
    Let $\A$ be an algebra that is, in particular, a lattice. For $S, T \in \term{X}{\A}$, let $=$ denote the syntactic equality on trees.
    A \textit{normal form function} is a function $F: \term{X}{\A} \to \term{X}{\A}$ satisfying the following three properties:

    \begin{align*}
    &\forall S,\ F(S) \simw{\A} S \\ 
    &\forall S, T,\ S \simw{\A} T \implies F(S) = F(T) \\
    &\forall S,\ F(S) \text{ is in minimal form (\autoref{def:minimalform})} 
    \end{align*}
\end{definition}

The first step is to construct such a function for $\BLP$.
The result of \autoref{sec:generalizedwordproblem} gives us an inductive characterization of $\leq_\OLP$.
It will be useful to have a similar one for $\leq_\BLP$.

\begin{definition}[\CFPBL\coqref{CFBLPlus}]
    Let {\CFPBL} be the proof system of \autoref{fig:proofSystemLP}. Note that {\CFPBL} is the restriction of {\CFP} to sequents with exactly one left-annotated term and one right-annotated term. {\CFPBL} is also the extension of Whitman's characterization of free lattices\cite{freeseFreeLattices1995, whitmanFreeLattices1941} with monotonic functions.
\end{definition}

\begin{figure}[ht]
\begin{center}
    \begin{tabular}{c c}
    \ \\
        \multicolumn{2}{c}{            
            \AxiomC{}
            \RightLabel{ \scrule{ Hyp}}
            \UnaryInfC{$S \vdash_\BLP S$}
            \DisplayProof
        }\\[3ex]
                
        \AxiomC{}
        \RightLabel{ \scrule{ LeftBot}}
        \UnaryInfC{$\bot^L \vdash_\BLP \Delta$}
        \DisplayProof &
        \AxiomC{}
        \RightLabel{ \scrule{ RightTop}}
        \UnaryInfC{$\Gamma \vdash_\BLP \top^R$}
        \DisplayProof
        \\[3ex]
        
        \AxiomC{$S \vdash_\BLP \Delta$}
        \RightLabel{ \scrule{ LeftAnd}}
        \UnaryInfC{$S \land T \vdash_\BLP \Delta$}
        \DisplayProof &
        \AxiomC{$\Gamma \vdash_\BLP S$}
        \AxiomC{$\Gamma \vdash_\BLP T$}
        \RightLabel{ \scrule{ RightAnd}}
        \BinaryInfC{$\Gamma \vdash_\BLP S \land T$}
        \DisplayProof
        \\[3ex]

        \AxiomC{$S \vdash_\BLP \Delta$}
        \AxiomC{$T \vdash_\BLP \Delta$}
        \RightLabel{ \scrule{ LeftOr}}
        \BinaryInfC{$S \lor T \vdash_\BLP \Delta$}
        \DisplayProof &
        \AxiomC{$\Gamma \vdash_\BLP S$}
        \RightLabel{ \scrule{ RightOr}}
        \UnaryInfC{$\Gamma \vdash_\BLP S \lor T$}
        \DisplayProof
        \\[3ex]
        \multicolumn{2}{c}{
            \AxiomC{$S_{x_1} \vdash_\BLP T_{x_1}$}
            \AxiomC{$T_{x_1} \vdash_\BLP S_{x_1}$\quad ...}
            \AxiomC{$S_{y_1} \vdash_\BLP T_{y_1}$\quad ...}
            \AxiomC{$T_{z_1} \vdash_\BLP S_{z_1}$\quad ...}
            \RightLabel{\scrule{ $F$-rule}}
            \QuaternaryInfC{$F(S_{x_1}, ..., S_{y_1}, ..., S_{z_1}, ...) \vdash_\BLP F(T_{x_1}, ..., T_{y_1}, ..., T_{z_1}, ...)$}
            \DisplayProof 
        }\\[3ex]
\ \\

    \end{tabular}
    \end{center}
    \caption{Deduction rules of \CFPBL.}    
\label{fig:proofSystemLP}
\end{figure}

\begin{theorem}[Soundness and Completeness of \CFPBL\coqrefline{CFBLPlus\_soundness, CFBLPlus\_completeness}] 
    \label{thm:cfpbl_sound_complete}
    For $S, T \in \term{X}{\BLP}$, $S \leq_\BLP P$ if and only if $S \vdash_{\CFP}P$ is provable.
\end{theorem}
\begin{proof}
    As in \autoref{lem:cfp_sound_complete}; see also the formal proof in Rocq.
\end{proof}

We now state a useful inversion lemma for terms starting with a function symbol.

\begin{lemma}
\label{lem:finversion}
    Let $T \in \term{X}{\BLP}$ be in minimal form and suppose $T \simw{\BLP} F(\Sxyz)$, where as usual $F(\xyz)$ is invariant in $x_i$, covariant in $y_i$ and contravariant in $z_a$.
    Then $T = F(\Txyz)$ and $S_{x_i} \simw{\BLP} T_{x_i}$, \ $S_{y_j} \simw{\BLP} T_{y_j}$, \ $S_{z_k} \simw{\BLP} T_{z_k}$.
\end{lemma}
\begin{proof}
    Consider the shape of $T$. 
    \begin{itemize}
        \item For $T = x$, no rule of \CFPBL can conclude $F(\Sxyz) \leq_\BLP x$.
        \item Similarly, $F(\Sxyz) \leq_\BLP 0$ and $1 \leq_\BLP F(\Sxyz)$ cannot be deduced.
        \item For $T = T_1 \lor T_2$, the only rule that can conclude $F(\Sxyz) \leq_\BLP T_1 \lor T_2$ is \scrule{RightOr} and hence one of $T_i$ is such that $F(\Sxyz) \leq_\BLP T_i$. But from $T_1 \lor T_2 \leq_\BLP F(\Sxyz)$, we deduce $T_i \leq_\BLP F(\Sxyz)$ and hence $T_i \simw{\BLP} F(\Sxyz) \simw{\BLP} T$, contradicting the assumption that $T$ is minimal. 
    
        \item A dual argument similarly rules out $T = T_1 \land T_2$.
        
        \item For $T = g(\Sxyz)$, note that the only rules that can deduce $$F(\Sxyz) \leq_\BLP T$$ if $T$ has the shape of a function is the \scrule{$F$-rule} so $T =F(\Txyz)$ and 
    \begin{itemize}
        \item $S_{x_i} \simw{\BLP} T_{x_i}$
        \item $S_{y_j} \leq_\BLP T_{y_j}$
        \item $T_{z_k} \leq_\BLP S_{z_k}$.
    \end{itemize}
    Symmetrically,  $T \leq_\BLP F(\xyz)$ implies
        \begin{itemize}
        \item $y_j' \leq_\BLP y_j$
        \item $T_{y_j} \leq_\BLP S_{y_j}$
        \item $S_{z_k} \leq_\BLP T_{z_k}$.
    \end{itemize}
    So it follows that $S_{x_i} \simw{\BLP} T_{x_i}$, $S_{y_j} \simw{\BLP} T_{y_j}$ and $S_{z_k} \simw{\BLP} T_{z_k}$.
    \end{itemize}
\end{proof}

\begin{theorem}
\label{thm:lattice_norm_condition}
    A term $T \in \term{X}{\BLP}$ is in normal form if and only if one of the following holds:
    \begin{itemize}
        \item $T \in X$
        \item $T \in \lbrace \bot, \top \rbrace$
        \item $T = F(T_1, ..., T_n)$ and each $T_i$ is in normal form
        \item $T = T_1 \lor ... \lor T_n$, and all the following hold:
        \begin{itemize}
            \item each $T_i$ is in normal form
            \item $T_i \not\leq_\BLP T_j$ for all $i \neq j$
            \item if $T_i = T_{i1} \land ... \land T_{im}$ then for all $j$, $T_{ij}\not\leq_\BLP T$
        \end{itemize}
        \item $T = T_1 \land ... \land T_n$ and the dual of the conditions above hold.
    \end{itemize}
\end{theorem}
\begin{proof}
The forward direction is trivial.
For the backward direction, let $S$ be in minimal form and such that $S \simw{\BLP} T$. We show that $S = T$ by induction on the structure of $T$.
    \begin{itemize}
        \item If $T \in X \cup \lbrace \bot, \top \rbrace$ then trivially $S = T$.
        \item If $T = F(T_1, ..., T_n)$, by \autoref{lem:finversion}, $S =  F(S_1, ..., S_n)$ and $T_i \simw{\BLP} S_i$. Since by hypothesis $T_i$ is in normal form and $S_i$ is in minimal form, by induction $S_i = T_i$ for all $i$ and hence $S = T$.
        
        \item Suppose $T = T_1 \lor ... \lor T_n$. 
        \begin{itemize}
            \item If $S = x$ then $x \leq_\BLP T_1 \lor ... \lor T_n$ and hence necessarily $\forall i,\  x \leq_\BLP T_i$. 
        Dually $T_1 \lor ... \lor T_n \leq_\BLP x$ and hence using \CFPBL, there must be some $T_i$ is such that $T_i \leq_\BLP x$, so $T_i \simw{\BLP} x \simw{\BLP} T = T_1 \lor ... T_n$, and in particular for any $j$, $T_j \leq T_i$, contradicting assumption. 
            \item If $S = \bot$ then forall $i$, $T_i \simw{\BLP} \bot$, contradicting assumption. If $S = \top$, then looking at {\CFP} there is some $i$ such that $T_i = \top$, again contradicting assumptions.

            \item By \autoref{lem:finversion}, $S$ cannot start with a function symbol.
        
            \item Now suppose that $S = S_1 \land ... \land S_m$. The only rules that can deduce $S \leq T$ are \scrule{LeftOr} and \scrule{RightAnd}, and hence either
        $$
        (\exists i,\ S_i \leq T ) \text{ or } (\exists i,\ S \leq T_i )
        $$
        In the first case, since $S_i \leq_\BLP T \simw{\BLP} S$ and $S\leq_\BLP S_i$ we obtain $S_i \simw{\BLP} S$, contradicting the minimality of $S$. 
        In the second case we have $T_1 \lor ... T_n = T \simw{\BLP} S \leq_\BLP T_i$, and in particular for all $j$, $T_j \leq T_i$, contradicting the assumption.

            \item Hence, we necessarily have $S = S_1 \lor ... \lor S_m \simw{\BLP} T_1 \lor ... \lor T_n$, from which follows $\forall i,\  S_i \leq T$.
        
        Now we want to show that $\forall i\ \exists j,\  S_i \leq T_j$.
        If $S_i \in X$ or $S_i$ starts with a function symbol, we must have $\exists j,\ S_i \leq T_j$ by inspecting the {\CFPBL} proof of $S_i \leq T_1 \lor ... \lor T_n$. 
        
        On the other hand, if $S_i = S_{i1} \land ... \land S_{il}$ then we have 
        $$ S_{i1} \land ... \land S_{il} \leq T_1 \lor ... \lor T_n$$
        By inspecting \CFP, we necessarily have either $\exists j,\ S_i \leq T_j$ or $\exists j,\ S_{ij} \leq T_\BLP$, but the later contradicts minimality of $S$ since we would have
        $$S \leq_\BLP S_1 \lor ... \lor S_{ij} \lor... \lor S_m \leq_\BLP T \simw{\BLP} S $$
        and in particular $S \simw{\BLP} S_1 \lor ... \lor S_{ij} \lor... \lor S_m$.
        
        By a similar argument, $\forall j\  \exists i,\  S_i \leq T_j$ must hold as well. Since both are antichains, we it must hold that $n = m$ and, after reordering, $S_i \simw{\BLP} T_i$. Using induction, $S_i = T_i$, and hence $S = T$.
        
        \end{itemize}
        \item The $T = T_1 \land ... \land T_n$ case admits a dual proof.
        
    \end{itemize}
\end{proof}

\begin{corollary}\label{cor:lattice_subterm_cond}
    A term $S \in \term{X}{\BLP}$ is in normal form if and only if all the subterms $T$ of $S$ of the form $T = T_1 \lor ... \lor T_n$ satisfy the two properties:
    \begin{enumerate}
        \item $T_i \not\leq_\BLP T_j$ for all $i \neq j$
        \item if $T_i = T_{i1} \land ... \land T_{im}$ then for all $j$, $T_{ij}\not\leq_\BLP T$
    \end{enumerate}
    and dually for subterms of the form $T = T_1 \land ... \land T_n$.
\end{corollary}

%%%here

\begin{theorem}[Normal Form Function for \BLP]
    \label{thm:normal_form_lp}
    There exists an algorithm that takes  as input $S \in \term{X}{\BLP}$ a term of size $n$, runs in time $\mathcal O(n^2)$ and outputs a normal form for $S$.
\end{theorem}
\begin{proof}
    The existence of a normalization algorithm for $\BLP$ follows from \autoref{thm:lattice_norm_condition} and extends the normalization algorithm for lattices from \cite{guilloudFormulaNormalizationsVerification2023}. 
    Define $\zeta: \term{X}{\BLP} \rightarrow \term{X}{\BLP}$ as given by \autoref{alg:zeta}.
    Note that $\zeta$ is idempotent and that $\forall T \in \term{X}{\BLP},\ T \simw{\BLP} \zeta(T)$. Let $\term{X}{\BLP}^\zeta$ be the range of $\zeta$. Note also that all terms in $\term{X}{\BLP}^\zeta$ recursively satisfy the first property of \autoref{thm:lattice_norm_condition}.

\begin{algorithm}[hbt]
\DontPrintSemicolon
\caption{Algorithm for the $\zeta$ function}\label{alg:zeta}
    \procedure{zeta(T: Term): Term}{
        \Switch{T}{
            \Case{x $\in X$}{\Return x}
            \Case{F(T$_1$, ..., T$_n$)}{\Return F(zeta(T$_1$)), ..., F(zeta(T$_n$))}
            \Case{T$_1$ $\lor$ ... $\lor$ T$_m$}{
                accepted $\gets$ List.empty\;
                \For{$1\leq i\leq m$}{
                    T$'_i$ $\gets$ zeta(T$_i$)
                }
                \For{$1\leq i\leq m$}{
                    \Switch{T$'_i$}{
                        \Case{T$'_{i1}$ $\land$ ... $\land$ T$'_{in}$}{
                            \If{there exists $j$ s.t. T$'_{ij} \leq_\BLP$ (T$'_1$ $\lor$ ... $\lor$ T$'_m$)}{
                                accepted.append(T$'_{ij}$)
                            }
                            \Else{accepted.append(T$'_i$)}
                        }
                        \Case{\_}{accepted.append(T$'_i$)}
                    }
                }
                \Return $\bigvee$(accepted)
            }
            \Case{T$_1$ $\land$ ... $\land$ T$_m$}{ dual to the case above}
        }
    }
\end{algorithm}

\iffalse
    
    $$
    \zeta(x) = x
    $$
    $$
    \zeta(F(a_1, ..., a_m)) = F(\zeta(a_1), ..., \zeta(a_m))
    $$
    $$
    \zeta(a_1 \lor ... \lor a_m) = \begin{cases}
        \zeta(a_1 \lor ... \lor a_{ij} \lor ... \lor a_m)  &\text{ if } a_i = (a_{i1} \land ... \land a_{in}) \\
                        & \text{ and } a_{ij} \leq_\BLP a_1 \lor ... \lor a_m\\
        \zeta(a_1) \lor ... \lor \zeta(a_m)  & \text{ otherwise}
    \end{cases}
    $$
    \fi

Now, define $\eta: \term{X}{\BLP}^\zeta \rightarrow \term{X}{\BLP}^\zeta$ according to \autoref{alg:eta}.

\begin{algorithm}[hbt]
\DontPrintSemicolon
\caption{Algorithm for the $\eta$ function}\label{alg:eta}
    \procedure{eta(T: Term): Term}{
        \Switch{T}{
            \Case{x $\in X$}{\Return x}
            \Case{F(T$_1$, ..., T$_n$)}{\Return F(eta(T$_1$)), ..., F(eta(T$_n$))}
            \Case{T$_1$ $\lor$ ... $\lor$ T$_m$}{
                \For{$1\leq i\leq m$}{
                    T$'_i$ $\gets$ eta(T$_i$)
                }
                antichain $\gets$ List.empty\;
                \For{$1\leq i\leq m$}{
                    \If{there exists $j$ s.t. T$'_i$ $\leq_\BLP$ T$'_j$ \&\& (!(T$'_j$ $\leq_\BLP$ T$'_i$) || $i$>$j$)}{
                        continue
                    }
                    \Else{antichain.add(T$'_i$)}
                }
                \Return $\bigvee$(antichain)
            }
            \Case{T$_1$ $\land$ ... $\land$ T$_m$}{ dual to the case above}
        }
    }
\end{algorithm}

\iffalse
$$
    \eta(x) = x
$$
$$
    \eta(F(a_1, ..., a_m)) = F(\eta(a_1), ..., \eta(a_m))
$$
$$
    \eta(a_1 \lor ... \lor a_m) = \begin{cases}
    \eta(a_1 \lor ... \lor a_{i-1} \lor a_{i+1} \lor ... \lor a_m)  & \text{ if } a_i \leq_\BLP a_j, \text{ $i \neq j$ } \\
    \eta(a_1) \lor ... \lor \eta(a_m) & \text{ otherwise}
\end{cases}
$$
\fi
Observe again that $\eta$ is idempotent and $\forall T \in \term{X}{\BLP},\ T \simw{\BLP} \eta(T)$, and let $\term{X}{\BLP}^{\zeta\eta}$ be the range of $\eta$. Since $\eta$ ensures that conjunctions and disjunctions are antichains, all elements of $\term{X}{\BLP}^{\zeta\eta}$ satisfy the conditions of \autoref{cor:lattice_subterm_cond} and hence are in normal form.

Now let $S, T \in \term{X}{\BLP}$ such that $S \simw{\BLP} T$. Then
$$\eta(\zeta(S)) \simw{\BLP} S \simw{\BLP} T \simw{\BLP} \eta(\zeta(T))$$ 
so $\eta(\zeta(S)) = \eta(\zeta(T))$.

Hence $\eta \circ \zeta$ is a normal form function. To analyze the runtime, observe as that $\eta$ and $\zeta$ each takes at most quadratic time to compute, plus the time required to compute $\leq_\BLP$. Note also that $\leq_\BLP$ is only ever necessary to compute at most on every pair of subterms of the normal form of the input, so that the algorithm of \autoref{sec:generalizedwordproblem} computes all of them in time $\O(n^2)$.
\end{proof}

\subsection{Normalization for \OLP}
The first technical difficulty for normalization in {\OLP} relates to negation normal form. Indeed, for ortholattices without function symbols we can define a function
$$
\delta: \term{X}{\OL} \mapsto \T_\L(X \cup X')
$$
where $X' = \lbrace \neg x \mid x \in X \rbrace$, corresponding to negation normal form such that $\eqclass{\delta(\cdot)}{\OL} : \term{X}{\OLP}_{\slashsimw{\OL}} \mapsto \T_\L(X \cup X')_\slashsimw{\OL}$ is an isomorphism of ortholattices. This is not immediately possible in $\OLP$, as negation cannot ``go through'' function symbols. Nonetheless, we can define $\delta: \term{X}{\OLP} \mapsto \term{X}{\OLP}$ as a pseudo-negation-normal form, that is:
\begin{center}
    
\begin{tabular}{r c l r c l}
     $\delta(x) $&$=$&$ x$ & $\delta(\neg x) $&$=$&$ \neg x$ \\
     $\delta(T_1 \lor t_2) $&$=$&$ \delta(T_1) \lor \delta(t_2)$ &  $\delta(\neg (T_1 \lor t_2)) $&$=$&$ \delta(\neg T_1) \land \delta(\neg t_2)$\\
     $\delta(T_1 \land t_2) $&$=$&$ \delta(T_1) \land \delta(t_2)$ &  $\delta(\neg (T_1 \land t_2)) $&$=$&$ \delta(\neg T_1) \lor \delta(\neg t_2)$\\
     $\delta(F(T_1, ...)) $&$=$&$ F(\delta(T_1), ...) $ & $\delta(\neg F(T_1, ...)) $&$=$&$ \neg F(\delta(T_1), ...) $\\
      & & & $\delta(\neg \neg T) $&$=$&$ \delta(T)$
\end{tabular}
\end{center}

Note that $\delta$ is idempotent, and let $\term{X}{\OLP}^\delta$ be the range of $\delta$, i.e. the terms in pseudo-negation-normal form.

\iffalse
\begin{lemma}
    A statement in $\term{X}{\OLP}^\delta$ has a proof in $\CFPD$ if and only if it has a proof in $\CFP$
\end{lemma}
\begin{proof}
    For the $\Longrightarrow$ direction, the \scrule{ NegSwap} rule can be simulated with one instance of the \scrule{LeftNeg} rule and one of the \scrule{RightNeg} rule. For the $\Longleftarrow$ direction, since negations only exist right above functions symbols and variables, applications of \scrule{RightNeg} and \scrule{LeftNeg} trivially swap up with every other rule, until they can be replaced by an appication of \scrule{NegSwap}.
\end{proof}
\fi

We now expand the signature of {\OLP} to \lOLP.

\begin{definition}
    Define {\lOLP} (resp \lBLP) as the algebraic signature of {\OLP} (resp \BLP) extended with, for each symbol $F$ in \OLP, a new symbol $\bar F$ with the opposite monotonicity.
    Note that $\term{X}{\OLP}^\delta$ and $\term{X\cup X'}{\lBLP}$ are isomorphic.
    To reduce notational burden, keep this isomorphism implicit and, with a slight abuse of notation, identify the expressions $\neg (F(\xyz)))$ and $\bar F(\xyz)$.
    
\end{definition}

\begin{definition}
    We define a new proof system $\CFPD$ as the system made of all the rules of $\CFP$ except for the two negation rules.
\end{definition}
\begin{lemma}\label{lem:proofindelta}
    For all $S, T \in \term{X\cup X'}{\lBLP}$, $\vdash_{\CFP} S^L, T^R \iff \vdash_{\CFPD} S^L, T^R$ .
\end{lemma}
\begin{proof}
    Forward: by induction on the proof of $\vdash_{\CFP} S^L, T^R$. Not that all proof steps but \scrule{LeftNot} and \scrule{RightNot} over functions can be directly replicated in $\CFPD$. Then, since negation only occurs right above function symbols and variables, \scrule{LeftNot} and \scrule{RightNot} steps can only occur in the following configuration (the case where the negation appear above a variable is similar):
    \begin{center}
    \AxiomC{$\A$}
    \UnaryInfC{$F(\Sxyz)^R, F(\Txyz)^L$}
    \RightLabel{\scrule{ LeftNot}}
    \UnaryInfC{$F(\Sxyz)^R, \neg F(\Txyz)^R$}
    \RightLabel{\scrule{ RightNot}}
    \UnaryInfC{$\neg F(\Sxyz)^L, \neg F(\Txyz)^R$}
    \DisplayProof 
\end{center}
    note that $\A$ is necessarily an instance of the \scrule{$F$-rule}, and hence the three steps can be replaced by one instance of the \scrule{ $\bar F$-rule}.

    Backward: by induction on the proof of $T^L, S^R$, the transformation inverse from the one above yields the desired proof of $S^L, T^R$
\end{proof}
\
The lemmas above allow us to retrieve a nnf-like form where negations only appear above literals and function symbols. 
We now define one last reduction function.
Let $\beta : \term{X\cup X'}{\lBLP} \rightarrow R$ according to \autoref{alg:beta}.

\iffalse
Let $R \subset \term{X}{\lBLP}^\delta$ be defined as follows:
    \[\begin{array}{ll}
&0, 1, x, x' \in R\ \ (\text{for }x \in X) \\
&F(\xyz) \in R \iff \xyz \in R \\
&a \lor b \in R  \iff  a\in R, b\in R,\ 
 \delta(\neg a) \nle_\BLP a \lor b,\ 
 \delta(\neg b) \nle_\BLP a \lor b\\

&a \land b \in R  \iff a\in R, b\in R,\ 
 \delta(\neg a) \nge_\BLP a \land b,\ 
 \delta(\neg b) \nge_\BLP a \land b

\end{array}\]

Above, $\leq_L$ is the order relation on lattices, $x \ge_\BLP y$ denotes
$y \le_\BLP x$, 
and $\nle_\BLP$, $\nge_L$ are the negations of those conditions: 
$x \nle_\BLP y$ iff not $x \le_\BLP y$, whereas
$x \nge_\BLP y$ iff not $y \le_\BLP x$.

\[\begin{array}{ll}
&\beta(0) = 0, \beta(1) = 1, \beta(x) = x, \beta(x') = x' \ (\text{for }x \in X) \\
&\beta(F(\xyz)) = F(\beta(x_1), ..., \beta(y_1), ..., \beta(z_1), ...)\\
&\beta(a \lor b) = \begin{cases}
\beta(a) \lor \beta(b) & \text{ if } \beta(a) \lor \beta(b) \in R\\
1 & \text{ otherwise} 
\end{cases} \\
&\beta(a \land b) = \begin{cases}
\beta(a) \land \beta(b) & \text{ if } \beta(a) \land \beta(b) \in R\\
0 & \text{ otherwise}
\end{cases}
\end{array}\]
\fi

\begin{algorithm}[hbt]
\DontPrintSemicolon
\caption{Algorithm for the $\beta$ function}\label{alg:beta}
    \procedure{beta(T: Term): Term}{
        \Switch{T}{
            \Case{x $\in X$}{\Return x}
            \Case{x' $\in X'$}{\Return x'}
            \Case{F(T$_1$, ..., T$_n$)}{\Return F(beta(T$_1$)), ..., F(beta(T$_n$))}
            \Case{T$_1$ $\lor$ T$_2$}{
                bt1 $\gets$ beta(T$_1$)\;
                bt2 $\gets$ beta(T$_2$)\;
            
                \If{(delta($\neg$ bt1) $\leq_\BLP$ bt1 $\lor$ bt2) | (delta($\neg$ bt2) $\leq_\BLP$ bt1 $\lor$ bt2)}{
                    \Return 1
                }
                \Else{
                    \Return bt1 $\lor$ bt2
                }
            }
            \Case{T$_1$ $\land$ T$_2$}{ dual to the case above}
        }
    }
\end{algorithm}
It is clear that $\forall S,\ \beta(S) \simw{\OLP} S$, and hence in particular $S \leq_{OL^+} T \iff \beta(S) \leq_{OL^+} \beta(T)$. Let $\term{X\cup X'}{\lBLP}^\beta$ be the range of $\beta$.

\begin{lemma}\label{lem:beta_cancel}
    If $S \in \term{X\cup X'}{\lBLP}^\beta$, then
    \begin{itemize}
        \item $S \simw{\OLP} 1 \iff phi \simw{\BLP} 1$
        \item $S \simw{\OLP} 0 \iff phi \simw{\BLP} 0$
    \end{itemize}
\end{lemma}
\begin{proof}
    By induction on $S$. Assuming $S \simw{\OLP} 1$, $S$ can only be $1$ (in which case the conclusion is immediate), a conjunction or a disjunction.
    If $S = \phi_1 \land \phi_2 \simw{\OLP} 1$ then both $\phi_1 \simw{\OLP} 1$ and $\phi_2 \simw{\OLP} 1$, so by induction $\phi_1 \simw{\BLP} 1$ and $\phi_2 \simw{\BLP} 1$ and the conclusion follows.
    If $S = \phi_1 \lor \phi_2 \simw{\OLP} 1$ then $S = 1$ by definition of $\beta$.

    The case $S \simw{\OLP} 0$ is, as always, dual.
\end{proof}

We are now left with showing that $\term{X\cup X'}{\lBLP}^\beta$ is an $\OL^+$ \textit{when seen with the rules of $\CFPBL$}.

\begin{lemma}
    \label{lem:scp_to_scpl}
For all $S$ and $T \in \term{X\cup X'}{\lBLP}^\beta$,
    $$\vdash_{\CFP} S^L, T^R \iff S \vdash_{\CFPBL}  T$$
\end{lemma}
\begin{proof}
    By \autoref{lem:proofindelta}, we only need to show $\vdash_{\CFPD} S^L, T^R \iff S \vdash_\CFPBL T$.
    The backward direction is trivial, as $\CFPBL$ is a subsystem of $\CFPD$. 
    For the forward direction, the proof proceeds by induction first on the number of instances of the \scrule{Replace} rule in the proof, then on the size of the proof tree of $S \leq_{\OL^{+\delta}} T$.
    
    The rules \scrule{Hyp}, \scrule{LeftZero}, \scrule{RightOne}, \scrule{LeftOr}, \scrule{RightOr}, \scrule{LeftAnd} and \scrule{RightAnd} of $\CFPD$ also exist in $\CFPBL$.The last case is \scrule{Replace}. Suppose that the replaced term is $S^L$ (A dual argument handles the case of aright term.):
    \begin{center}
        \AxiomC{$\A$}
        \UnaryInfC{$S^L, S^L$}
        \RightLabel{\scrule{ \scrule{Replace}}}
        \UnaryInfC{$S^L, \Gamma$}
        \DisplayProof 
    \end{center}
    Note that if $\vdash_{\CFPD} S^L, S^L$ holds, then $S \simw{\OLP} 0$. Hence, by \autoref{lem:beta_cancel}, $S \simw{\BLP} 0$. Then by completeness (\autoref{thm:cfpbl_sound_complete}) there exists a proof of $S \vdash_{\CFPBL} \delta(\neg S)$.
    
    \iffalse
    can be straightforwardly translated dually to a proof $\A'$ of $\vdash_{\CFPD} S^L, \delta(\neg S)^R$ so that in particular the number of instances of \scrule{Replace} in $\A'$ is the same as in $\A$ and they have the same size. Hence by induction there exists a proof of $S \vdash_{\CFPBL} \delta(\neg S)$.
    However since $S \in \term{X\cup X'}{\lBLP}^\beta$, by definition we must have $S = 0$, and $0^L, \Gamma$ is trivially provable in $SC^{L^+}$.
    \fi
\end{proof}

\begin{corollary}
    \ \ $\eqclass{\term{X\cup X'}{\lBLP}^\beta}{\dashvdash_{LP}}$ is isomorphic to the free \OLP.
\end{corollary}

\begin{theorem}[Normalization for \OLP]
    {\OLP} admits a normalization function computable in quadratic time.
\end{theorem}

\begin{proof}
This normalization function is the composition of all the individual simplification steps, mapping $S \in \term{X}{\OLP}$ to $\eta(\zeta(\beta(\delta(S))))$. Indeed, let $S, T \in \term{X}{\OLP}$ such that $S \simw{\OLP} T$. Then,
$$
\beta(\delta(S)) \simw{\OLP} \beta(\delta(T))
$$
which by \autoref{lem:scp_to_scpl} implies
$$
\beta(\delta(S)) \simw{\BLP} \beta(\delta(T))
$$

And then by \autoref{thm:normal_form_lp} this implies
$$
\eta(\zeta(\beta(\delta(S)))) = \eta(\zeta(\beta(\delta(T))))
$$

Regarding the time complexity, note that $\delta$ can be computed in linear time and that $\beta$ takes quadratic time plus the time needed to compute $\leq_\BLP$. Note that all formulas that appear during the evaluation of $\eta(\zeta(\beta(\delta(S))))$, are (the normal forms of) subformulas of $S$ or negation of such formulas, so that we indeed need only quadratic time to compute the $\leq_\BLP$ relation on all the necessary pairs. 
\end{proof}

%\section{Type Checking}
%\label{sec:typechecking}
%\input{sections/typechecking}

%\section{Record Types}
%\input{sections/records}

%\section{Classes, Inheritance and Declarative Subtyping}
%\input{sections/classes}

\section{Related Work}

\label{sec:relatedwork}

Subtyping with union, intersection, negation, monotonic constructors and constraints have received considerable attention over the years \cite{DBLP:journals/iandc/MacQueenPS86, DBLP:journals/ndjfl/CoppoD80, DBLP:conf/popl/DunfieldP04, jiangBidirectionalHigherRankPolymorphism2025,jiangBidirectionalHigherRankPolymorphism2025}.
Whereas these systems admittedly support features beyond the scope of our work, many of them have high complexity of type checking or non-intuitive rules and restrictions. We believe that the key concept that distinguish our approach and yield a simpler, efficient and general system is the presentation of the problem in terms of provability of inequalities in a free (ortho)lattice.

Similarly, multiple work from around the start of the 21st century establish decidability and hardness results (from coNP to undecidability) for variants of the entailment problem (and the special case of validity and satisfiability) on subtyping constraint \cite{pottierSimplifyingSubtypingConstraints1996, hengleinConstraintAutomataComplexity1998, suFirstorderTheorySubtyping2002, kuncakStructuralSubtypingNonrecursive2003} in a type lattice.  However, as discussed in Sections 1 and \ref{sec:typelattices}, they all consider the problem over the specific model of syntactic types built from $\to$ and a lattice of base types, assuming for example the equality $(A \to B) \land (C \to D) = (A\lor C)\to (C\land D)$, and even, for some of them, $(A, A) \lor (B, B) = (A\lor B, A\lor B)$, which would not be sound in more expressive languages, as discussed in \autoref{sec:typelattices}. This is different from our algorithm, which solves the problem on the free lattice. Not only our approach allows much faster algorithms, it also satisfies the open world assumption: any accepted subtyping relation remains true if more types or programming constructs are later introduced, which is not the case with a fixed syntactic model. Syntactic models with a mathematical mental model of types and functions.

More recently, \cite{dolanAlgebraicSubtypingDistinguished2017} presented a type inference algorithm for MLsub, an extension of ML with subtyping. The system uses union and intersection types internally but in a limited way, to express bounds on type variables. The problem is shown decidable, but no mention of complexity is made.

In \cite{parreauxMLstructPrincipalType2022}, MLsub is extended to MLstruct, where union and intersection are not limited. MLstruct also supports negation types. However, union and negation types do not conform to intuitive specification of expressive union types: For example, the type \lstinline+{x: A} | {y: B}+ is equivalent to the top type $\top$. 
Polymorphic inference of principal types in the system is proven decidable, without mention of complexity. MLstruct comprises the core features of the programming language MLscript \cite{lionelparreauxHkusttacoMlscript2025}.

Indicative of interest in negation types, 
\cite{castagnaProgrammingUnionIntersection2024} describes a semantic set-theoretic subtyping system used in the language CDuce~\cite{cduceCDuceCompiler2021}. The language assigns a set of terms to every type, where unions, intersection and negation are interpreted as their set-theoretic counterpart and subtyping reduces to inclusion. The language supports overloaded function, match patterns and bounded polymorphism, but not F-bounded polymorphism nor constraints, and impose strong assumptions and restrictions to obtain a decision procedure. The system and algorithms are impressive but rather complex, and no mention of complexity is made (\textit{at least} coNP).

\cite{jiangBidirectionalHigherRankPolymorphism2025} demonstrates an algorithm for partial type inference (of base types) in the presence of higher-rank polymorphism, unions and intersection types and explicit type application. Notably, their system $F^e_{\sqcup\sqcap}$ does \textit{not} assume distribution of intersection over function types, and is hence meaningfully able to use types such as $(Int \to Int) \land (Bool \to Bool)$. The system has been formalized but the computational complexity of type checking is not discussed.

Our algorithms can also be interpreted as decision procedures for certain classes of logical constraints. Related results include decidability of extensions of lattices with monotonic functions \cite{DBLP:conf/synasc/Sofronie-Stokkermans14} and subsequent complexity results based on an $O(n^4)$ normalization algorithms. These results are formulated in terms of algebraic constraints combined using classical propositional logic, which makes the overall problem NP hard. In the future, our proof system could perhaps be used to improve the behaviour of such algorithms on conjunctions of literals.

From a logic point of view, orthologic with constructors provides a framework for reasoning about approximations of classical logical constraints. In this context, ortholattice operations approximate Boolean algebra ones, whereas monotonic operators approximate universal and existential quantifiers. Such ideas have been explored (without analysis of their completeness) in the Lisa proof assistant \cite{guilloudLISAModernProof2023}. Aside from applications to executable languages, our work could also be used to introduce a richer soft type system in set theoretic systems such as Lisa, where an HOL type system was demonstrated \cite{guilloudMechanizedHOLReasoning2024}.

While ortholattices have been described by mathematicians in the last century, their proof theory and applications in computer science have been subject to more recent research interest
\cite{laurentFocusingOrthologic2016,kawanoLabeledSequentCalculus2018,guilloudFormulaNormalizationsVerification2023,guilloudInterpolationQuantifiersOrtholattices2024,guilloudOrthologicAxioms2024}. In particular, they have been used as an efficient approximation to classical logic in verification tools. However, monotonic function symbols in the language have never been considered, and we expect that our result will be similarly applicable to verification.

\section{Conclusions}
We have presented a new approach to subtyping in the presence of union, intersection and negation types, covariant, contravariant and invariant type constructors and subtyping constraints. 
Despite the simplicity of the formulation of our problem, the existing solutions face challenges including high complexity and subtlety of syntactic rules, inefficiency, incompleteness, or even unsoundness. 

Our approach is based on the provability of subtyping inequalities when the class of all types is considered to form an ortholattice with arbitrary function symbols (\OLP). This makes the approach open-world-compatible and applicable to many type systems and models of types. Moreover, the support for type constructors and arbitrary constraints allow us to directly express many common as well as uncommon features of modern type systems, such as constrained polymorphism and nominal subtyping declarations.

The main appeal of the approach is the existence of efficient algorithms for subtyping and type normalization. We have demonstrated, using proof theory of orthologic, that \OLP with assumptions admits a sound, complete and algorithmic set of rules. We have formalized this key result in the Rocq proof assistant (enclosed in the supplementary material).
The resulting algorithm allows deciding the entailment problem for subtyping relations. It is compatible with structure sharing and can solve multiple queries simultaneously, making it practical.
Additionally, we showed that \OLP admits a quadratic time normal form function, mapping each terms to the unique term of smallest size in its equivalence class. This allows to reliably simplify types in a compiler.

Thanks to our use of ortholattices, our approach yields principled algorithms for systems with negation, accounting for rules such as de Morgan laws. This is of interest for many recent expressive type systems. That said, even when the types do not refer to the negation operator, our results yield corresponding theorems and algorithms for the theory of bounded lattices with function symbols (which can be monotonic, antinomotonic, or unconstrained). We expect these to be useful in its own right, even for the systems that do not contain a negation operator on types.

\raggedright
\bibliographystyle{ACM-Reference-Format}
\bibliography{sguilloud,more}

\newpage

\appendix
\appendix
\section{Additional Lemmas And Proofs}
\label{apdx:proofs}

\begin{lemma}[Substitution]
\label{lem:substitution}
Let $F(A_1, ..., A_n)$ be a function symbol of arity $n$, and let $G$ be an arbitrary term such that for every $A_i$, if $F$ is covariant (respectively contravariant) in $A_i$ then $G$ is covariant (respectively contravariant) in $A_i$. Then for any set of axiom $C$ and annotated terms $\Gamma$ and $\Delta$ such that
$$\vdash_C \Gamma, \Delta$$
We have
$$\vdash_{C[F := G]} \Gamma[F := G], \Delta[F := G]$$
\end{lemma}
\begin{proof}
    Follows from soundness. Since $\vdash_C \Gamma, \Delta$ is provable, it necessecarily holds in every \OLP with any valuation of function symbols and variables that satisfies axioms in C. One particular such \OLP is $\T_\OLP$ with the valuation map that is the identity on every variable and every function symbol but $F$, mapping F to the functions that maps $x_1, ..., x_n$ to $G[A_1 := x_1, ... A_n := x_n]$.
    Note that we require the term $G$ to have the same variance as $F$, as otherwise it would not satisfy V10.
\end{proof}

\begin{definition}[Abstract Type Definition]
    Let a \textit{type definition} for a context $C, \Gamma$ be of the form
    \begin{center}
        \lstinline|type |$T[A_1, A_2, ...] <: F_{A_1, A_2, ...}$
    \end{center}
    Where $T$ is a new constant type that does not appear in T, $A_i$ are type variables assumed to be fresh for $C$ and \lstinline|F|$_{A_1, A_2, ...}$ is an arbitrary type whose variables are among the $A_i$. If $F_{A_1, A_2, ...}$
\end{definition}
The example above corresponds to the three type definitions 
\begin{lstlisting}
        type $S[A] <: \top$
        type $T[A] <: \top$
        type $U[A] <: S[A] \land T[S[A]]$
\end{lstlisting}
\begin{theorem}
    \label{thm:abstracttypesdefinitions}
    Given a context $C$ and an abstract type definition $T[A_1, A_2, ...] <: F_{A_1, A_2,...}$, let $C' :=  C \cup \lbrace T[A_1, A_2, ...] <: F_{A_1, A_2,...} \mid A_1, A_2, ... \in \term{X}{\OLP}\rbrace$.
    There exists a type $G[A_1, A_2, ...]$ such that for any two types $S_1$ and $S_2$, 
        $$S_1 <:_{C'} S_2$$
    if and only if
        $$S_1[T := G] <:_C S_2[T := G]$$

\end{theorem}

\begin{proof}
    Define $G[A_1, A_2, ...] := F_{A_1, A_2,...} \land T'[A_1, A_2, ...]$. Let 
    $$C' :=  \cup \forall A_1, A_2, ... . T[A_1, A_2, ...] <: F_{A_1, A_2,...}$$
    \begin{itemize}
        \item $(\Longrightarrow)$ Assume $S_1 <:_{C'} S_2$. By \autoref{lem:substitution}, we have
        $$S_1[T := G] <:_{C'[T := G]} S_2[T := G]$$

        Note that for any particular instantiation of the $A_i$'s, 
        $$G_{A_1, A_2, ...} = F_{A_1, A_2,...} \land T'[A_1, A_2, ...] <: F_{A_1, A_2,...}$$
        so the axioms are redundant.  Hence
        $$S_1[T := G] <:_{C} S_2[T := G]$$

        \item $(\Longleftarrow)$ Assume $S_1[T := G] <:_{C} S_2[T := G]$. Since $C \subseteq C'$, we have
        $$S_1[T := G] <:_{C'} S_2[T := G]$$
        By \autoref{lem:substitution} and since $C'$ does not contain $T'$, we have that 
        $$S_1[T := G][T' := T] <:_{C'} S_2[T := G][T' := T]$$
        Since $T'$ only appears inside of G, this is equal to
        $$S_1[T := G[T' := T]] <:_{C'} S_2[T := G[T' := T]]$$
        But note that 
        $$G[T' := T][A_1, A_2, ...] = F_{A_1, A_2,...} \land T[A_1, A_2, ...] =_{C'} T[A_1, A_2, ...]$$ using the absorption law.
        Hence we obtain $S_1 <:_{C'} S_2$
        
    \end{itemize}
\end{proof}

\end{document}